\documentclass[12pt]{article}
\pdfoutput=1
\usepackage{makeidx}
\makeindex
\usepackage[a4paper]{geometry}
\usepackage{jheppub,amsmath,amssymb,amsfonts,amsxtra,mathrsfs,graphics,graphicx,amsthm,epsfig,ytableau,bm,longtable,float,color,tikz,mathtools,xfrac,footnote,rotating,lscape}
\usepackage{dsfont}
\usepackage{multicol}
\usepackage{lipsum}
\usepackage{textgreek}
\usetikzlibrary{decorations.pathmorphing}
\usetikzlibrary{decorations.markings}
\usetikzlibrary{quotes,arrows.meta}
\usetikzlibrary{arrows, decorations.markings, calc, fadings, decorations.pathreplacing, patterns, decorations.pathmorphing, positioning}
\usepackage{tikz-cd}

\usepackage{fixmath} 
\usepackage{scalerel}
\newlength\bshft
\def\fakebold#1{\ThisStyle{\ooalign{$\SavedStyle#1$\cr%
  \kern-\bshft$\SavedStyle#1$\cr%
  \kern\bshft$\SavedStyle#1$}}}

\usetikzlibrary{positioning,shapes}
\usetikzlibrary{chains}
\usetikzlibrary{arrows,fit,decorations.pathreplacing}
\tikzstyle{every picture}+=[remember picture]
\tikzstyle{na} = [baseline=-.5ex]

\usepackage{empheq}
\usepackage{multirow}
\usepackage{booktabs}
\usepackage[american]{babel}

\usepackage[latin1]{inputenc}

\definecolor{cardinal}{rgb}{0.6,0,0}
\definecolor{darkgreen}{rgb}{0,0.4,0}
\definecolor{purple}{rgb}{0.5, 0, 0.5}
\definecolor{golden}{rgb}{0.92, 0.7, 0}
\definecolor{midnight}{rgb}{0, 0, 0.5}
\definecolor{darkblue}{rgb}{0, 0, 0.8}

\makeatletter
\newcommand{\vast}{\bBigg@{1}}
\newcommand{\Vast}{\bBigg@{5}}
\makeatother

\usepackage{hyperref}

\numberwithin{equation}{section}

\newcommand{\eg}{\textit{e.g.}}

\newcommand{\ie}{\textit{i.e.}}

\newcommand{\ii}{\mathrm{i}}
\newcommand{\?}{\;\!}

\numberwithin{equation}{section}

\newcommand{\be}{\begin{equation}} \newcommand{\ee}{\end{equation}}
\newcommand{\bea}{\begin{equation} \begin{aligned}} \newcommand{\eea}{\end{aligned} \end{equation}}

\newcommand{\Iprod}[2]{\langle {#1}, {#2} \rangle}

\def\U{\mathrm{U}}

\def\SU{\mathrm{SU}}

\def\u{\mathsf{u}}
\newcommand{\ex}{{\mathrm{e}}}
\newcommand{\rd}{\mathrm{d}}

\newcommand{\wt}{\widetilde}

\DeclareMathOperator{\Tr}{Tr}

\DeclareMathOperator{\sign}{sign}

\DeclareMathOperator{\re}{\mathbb{R}e}
\DeclareMathOperator{\im}{\mathbb{I}m}

\newcommand{\pd}{\partial}

\newcommand{\cF}{\mathcal{F}}

\newcommand{\cH}{\mathcal{H}}
\newcommand{\cI}{\mathcal{I}}
\newcommand{\cJ}{\mathcal{J}}

\newcommand{\cM}{\mathcal{M}}
\newcommand{\cN}{\mathcal{N}}
\newcommand{\cO}{\mathcal{O}}

\newcommand{\cQ}{\mathcal{Q}}

\newcommand{\cW}{\mathcal{W}}

\newcommand{\bR}{\mathbb{R}}
\newcommand{\bZ}{\mathbb{Z}}

\newcommand{\fg}{\mathfrak{g}}

\newcommand{\fh}{\mathfrak{h}}

\newcommand{\fm}{\mathfrak{m}}

\newcommand{\fn}{\mathfrak{n}}

\newcommand{\fR}{\mathfrak{R}}

\usepackage[Symbol]{upgreek}
\usepackage{bm}

\usepackage{calligra}
\DeclareMathAlphabet{\mathcalligra}{T1}{calligra}{m}{n}

\theoremstyle{plain}

  \theoremstyle{definition}

\providecommand{\examplename}{Example}
\providecommand{\theoremname}{Theorem}

\makeatletter
\g@addto@macro\bfseries{\boldmath}
\makeatother

\makeatletter
\newcommand*{\rom}[1]{\expandafter\@slowromancap\romannumeral #1@}
\makeatother

\title{Microstates of rotating AdS$_5$ strings}

\author[a]{Seyed Morteza Hosseini,}
\author[b]{Kiril Hristov,}
\author[c,d]{and Alberto Zaffaroni}
\affiliation[a]{Kavli IPMU (WPI), UTIAS, The University of Tokyo, Kashiwa, Chiba 277-8583, Japan}
\affiliation[b]{Institute for Nuclear Research and Nuclear Energy, Bulgarian Academy of Sciences, \\Tsarigradsko Chaussee 72, 1784 Sofia, Bulgaria}
\affiliation[c]{Dipartimento di Fisica, Universit\`a di Milano-Bicocca, I-20126 Milano, Italy}
\affiliation[d]{INFN, sezione di Milano - Bicocca, I-20126 Milano, Italy}
\emailAdd{morteza.hosseini@ipmu.jp}
\emailAdd{khristov@inrne.bas.bg}
\emailAdd{alberto.zaffaroni@mib.infn.it}

\preprint{IPMU19-0127}

\abstract{We provide a general formula for the refined topologically twisted index of $\mathcal{N}=1$ gauge theories living on the world-volume of D3-branes at conical Calabi-Yau singularities in the Cardy limit. The index is defined as  the partition function on $T^2 \times S^2_\omega$, with a partial topological twist and a $\Omega$-deformation along $S^2$, in the presence of background magnetic fluxes and fugacities for the global symmetries and can be used to study the properties of a class of BPS black strings. To this purpose, we find rotating domain-wall solutions of five-dimensional gauged supergravity interpolating between AdS$_5$ and a near horizon region consisting of a warped fibration of BTZ over a sphere. We explicitly construct rotating domain-walls that can be embedded in AdS$_5 \times S^5$ by uplifting a class of four-dimensional rotating black holes. We then provide a microscopic explanation of the entropy of such black holes by using  the refined topologically twisted index of $\mathcal{N}=4$ super Yang-Mills.}

\begin{document}

\setcounter{tocdepth}{2}
\maketitle

%
%

\date{Dated: \today}




\section{Introduction}
\label{sect:intro}

In this paper we consider theories living on D3-branes sitting at the tip of a Calabi-Yau three-fold cone and the relation of their twisted compactifications on a sphere to AdS black hole and black string physics. The holographic description of these twisted compactifications  consists of solutions interpolating between AdS$_5$ and AdS$_3\times S^2$ vacua. These can be interpreted as  renormalization group (RG) flows from an ultraviolet (UV) four-dimensional $\cN=1$ conformal field theory (CFT)  and an infrared (IR) two-dimensional $\cN = (0,2)$ one.  The right-moving central charge of the two-dimensional CFT has  been computed  in  \cite{Benini:2012cz,Benini:2013cda,Benini:2015bwz}, and successfully compared with the supergravity result for a variety of models.  In \cite{Hosseini:2016cyf} the partition function of the boundary theory on $T^2\times S^2$,  the so-called topologically twisted index  \cite{Nekrasov:2014xaa,Closset:2013sxa,Benini:2015noa,Honda:2015yha,Closset:2017bse}, was computed in the Cardy limit $\beta = -2 \pi \ii \tau \to 0$,  where $\tau$ is the modular parameter of the torus, and at finite $N$. It was found  that \cite{Hosseini:2016cyf}
\be
 \label{Cardylim}
 \log Z ( \fn , \Delta  | \beta )  = \frac{\pi^2}{6 \beta} c_l ( \fn , \Delta / \pi ) \, ,
\ee
where $c_l ( \fn , \Delta / \pi )$ is the left-moving trial central charge of the two-dimensional SCFT (see \eqref{chargesc}), $\fn$ denotes the set of magnetic fluxes of the twisted compactification and $\Delta$ denotes the set of  chemical potentials for the global symmetries of the theory. If we further compactify the black strings on a circle inside AdS$_3$ adding a momentum, we
obtain a four-dimensional static black hole with a hvLif asymptotic behaviour \cite{Hristov:2014eza}. As discussed in details in \cite{Hristov:2014eza,Hosseini:2016cyf,Hosseini:2018qsx,Zaffaroni:2019dhb}, a microscopic counting based  on \eqref{Cardylim} correctly reproduces the entropy of such black holes, which is just given by the Cardy formula in terms of the {\it exact} central charge $c_l^{\text{CFT}}  ( \fn  )$ of the two-dimensional SCFT.

It is the purpose of this paper to extend the previous picture to the rotating case, by computing the relevant quantum field theory partition function and by finding rotating black string solutions that can be embedded in consistent string compactifications.  

From the field theory point of view we need to compute the {\it refined} topologically twisted index \cite{Benini:2015noa}, which is defined as the partition function of an $\cN=1$ theory on   $T^2\times S^2_\omega$ with a topological $A$-twist  and an $\Omega$-background label by a complex parameter $\omega$ on $S^2$. It can be also written as the trace 
 \be
 Z (\fn, y , \zeta | q) =  \Tr_{\cH_{S^2_\omega \times S^1}}  (-1)^F q^{H_L} \zeta^{2 J} \prod_I y_I^{Q_I}  \, ,
\ee
where $q = \ex^{2 \pi \ii \tau}$, $y_I= \ex^{\ii \Delta_I}$, and  $\zeta = \ex^{\ii \omega / 2}$ are the fugacities associated to the left-moving Hamiltonian $H_L$, the flavor charges $Q_I$ and the angular momentum $J$ along $S^2_\omega$. We will show that, for a generic theory of D3-branes at toric (but not only) conical singularities, the Cardy limit   $\beta = -2 \pi \ii \tau\rightarrow  0$ (at finite $N$) of the topologically twisted index reads
\bea\label{cardylim2}
 \log Z ( \fn , \Delta , \omega  | \beta )  = \frac{\pi^2}{6 \beta} c_l ( \fn , \Delta / \pi )
 - \frac{( 2 \omega)^2}{27 \beta} \left( 3 c(\fn) - 2 a(\fn) \right) ,
\eea
where $c(\fn)$ and $a(\fn)$ are the trial central charges  (see \eqref{a:c:trial}) of the four-dimensional SCFT evaluated as functions of the magnetic fluxes. 

From the gravity point of view, we will find a new class of rotating black strings that can be embedded in AdS$_5\times S^5$. In five-dimensional language, these are domain walls  that interpolate between AdS$_5$ and a near horizon region consisting of a warped  fibration of BTZ over a sphere. To find these solutions, it is convenient to dimensionally reduce them to four dimensions, as also suggested by the Cardy limit we are performing in field theory, and construct the corresponding rotating asymptotically hvLif black holes. Luckily, a large class of dyonic  rotating black holes in four-dimensional $\cN=2$ gauged supergravity have been found recently in \cite{Hristov:2018spe} and we will use these results. Using the the 5D/4D relation, we can first find  the solutions in the so-called STU model in  four-dimensional $\cN=2$ gauged supergravity coupled to vector multiplets, then uplift them to a five-dimensional gauged supergravity, which is a consistent truncation of type IIB on AdS$_5\times S^5$, and, finally we could uplift our solutions to type IIB. We will  present the  general class of solutions for the STU model and, more generally, for symmetric models of gauged supergravity with vector multiplets. Finding solutions in consistent truncations of type IIB compactifications on more general Sasaki-Einstein
manifolds is more complicated and we leave it for future work.
 
We will then show that, for the twisted compactification of $\cN=4$ super Yang-Mills (SYM), the Legendre transform of   \eqref{cardylim2} exactly reproduces the entropy of the dyonic rotating black holes  that can be embedded in AdS$_5\times S^5$. This is another instance of the $\cI$\emph{-extremization} principle introduced in \cite{Benini:2015eyy,Benini:2016rke} and that has been successfully used to give a microscopic explanation of the entropy of BPS black holes in diverse dimensions.

In this context, it is also interesting to rewrite  the large $N$ limit of \eqref{cardylim2} as 
\be\label{exp}
 \log Z (\fn, \Delta , \omega | \beta) = \ii \sum_I \fn_I \frac{\partial \wt \cW (\Delta|\beta)}{\partial \Delta_I}
 + \frac{\ii \omega^2}{24} \sum_{I , J , K} \fn_I \fn_J \fn_K \frac{\partial^3 \wt \cW (\Delta|\beta)}{\partial \Delta_I \partial \Delta_J \partial \Delta_K} \, ,
\ee
where
\be
 \wt \cW (\Delta|\beta) = \frac{16 \pi^3 \ii}{27 \beta} a ( \Delta / \pi) \, ,
\ee
is the on-shell value of the twisted superpotential evaluated on the Bethe vacuum  \cite{Nekrasov:2014xaa,Benini:2015eyy,Closset:2017bse}  that dominates the index in the large $N$ limit. The first term in \eqref{exp} was already derived in \cite{Hosseini:2016cyf}. It appears in the same form in the expression of the three-dimensional topologically twisted index at large $N$ \cite{Benini:2015eyy,Hosseini:2016tor,Hosseini:2016ume}. It has been called {\it index theorem} and it is the field theory counterpart of the attractor mechanism of gauged supergravity \cite{Ferrara:1995ih,Cacciatori:2009iz,DallAgata:2010ejj}. It also appears, in a similar form,  in the five-dimensional topologically twisted index at large $N$ \cite{Hosseini:2018uzp}.
 
The paper is organized as follows. In section  \ref{sect:RTTI} we review the definition of the refined topologically twisted index on $T^2\times S^2_\omega$. In section \ref{sect:Cardy} we analyse the index in the Cardy limit and derive \eqref{cardylim2}. In section \ref{sect:rotating:strings} we use the 4D/5D connection to explicitly construct black string solutions that can be embedded in AdS$_5\times S^5$ and we compute the entropy of the corresponding four-dimensional black holes. In section  \ref{sect:holography} we compare the field theory and supergravity result finding complete agreement. We conclude in section \ref{sect:discussion} with discussions and open problems.
 
\section{The refined topologically twisted index}
\label{sect:RTTI}

Consider an $\cN = 1$ gauge theory with vector and  chiral multiplets in a representation $\oplus_I\fR_I$ of the gauge group $G$, and a non-anomalous $\U(1)_R$ symmetry in four dimensions.
The topologically twisted index for this class of gauge theories is defined as the (Euclidean) partition function on $T^2 \times \Sigma_\fg$, with a partial topological $A$-twist along the genus $\fg$ Riemann surface $\Sigma_\fg$ \cite{Benini:2015noa}.
It depends on the complex structure of the torus $q=\ex^{2 \pi \ii \tau}$, fugacities $y_I = \ex^{\ii \Delta_I}$ and magnetic fluxes $\fn_I$ (that are parameterizing the twist) for the global symmetries of the theory.
In the case of $\fg = 0$ one can refine the index by the angular momentum on $S^2$ and introduce the fugacity $\zeta = \ex^{\ii \omega / 2}$.
The index can be computed using supersymmetric localization and it is given by a matrix integral over the zero mode gauge variables $x = \ex^{i u}$ parameterizing the Wilson lines on the two directions of the torus
\be
 u = 2 \pi \oint_{\textmd{A-cycle}} A - 2 \pi \tau \oint_{\textmd{B-cycle}} A \, , 
\ee
which are defined modulo 
\be
 u_i \sim u_i + 2 \pi n + 2 \pi m \tau\, ,\qquad\qquad  n\, ,m \in \bZ \, .
\ee
The result is summed over a lattice of gauge magnetic fluxes $\fm$ (up to gauge transformations) on $\Sigma_\fg$ living in the co-root lattice $\Gamma_{\fh}$ of the gauge group.

The \emph{refined} topologically twisted index is thus explicitly given by a contour integral of a meromorphic differential form \cite{Benini:2015noa}%
\footnote{Supersymmetric localization selects a particular contour of integration and the final result can be cast in terms of the Jeffrey-Kirwan residue \cite{Benini:2015noa}.}
\bea
 \label{main:RTTI}
 Z ( \fm , x ; \fn , y , \zeta | q) = \frac{1}{|\mathfrak{W}|} \sum_{\fm \in \Gamma_\fh} \int_{\text{JK}} & \bigg( \prod_{\text{Cartan}} \frac{\rd x}{2 \pi \ii x} \eta(q)^2 \bigg)
 (-1)^{\sum_{\alpha > 0}\alpha (\fm)} \prod_{\alpha \in G}
 \bigg( \frac{\theta_1(x^\alpha \zeta^{|\alpha(\fm)|} ; q)}{\ii \eta(q)} \bigg) \\
 & \times \prod_{I} \prod_{\rho_I \in \fR_I} \prod_{j = - \frac{|B_I| - 1}{2}}^{\frac{|B_I| - 1}{2}}
 \bigg( \frac{\ii \eta(q)}{\theta_1(x^{\rho_I} y^{\nu_I} \zeta^{2j} ; q)} \bigg)^{\sign(B_I)} \, ,
\eea
summed over $\fm \in \Gamma_\fh$, where $B_I = \rho_I (\fm) - \nu_I ( \fn ) + 1$ and $|\mathfrak{W}|$ is the order of the Weyl group of $G$.
Here, $\alpha$ denotes the roots of $G$, and $\rho_I$, $\nu_I$ are the weights of the chiral multiplets under the gauge and flavor symmetry group, respectively.
Moreover, $\eta(q)$ is the Dedekind eta function and $\theta_1(x; q)$ is a Jacobi theta function (see appendix \ref{app:special:functions}).
Given the vanishing of the gauge and the gauge-flavor anomalies, the integrand in \eqref{main:RTTI} is a well-defined meromorphic function on the torus.

Finally, the invariance of the superpotential of the theory $W = \sum_a W_a$ under global symmetries imposes the constraint
\be\label{constrDelta}
 \prod_{I \in W_a} y_I = 1 \, , \quad \text {or } \sum_{I \in W_a} \Delta_I \in 2 \pi \bZ \, ,
\ee
where the product and the sum are restricted to the fields entering in the monomial $W_a$.
We have a similar constraint on flavor magnetic fluxes
\be\label{Rfluxes}
 \sum_{I \in W_a} \fn_I = 2 \, ,
\ee
that we call the \emph{twisting condition} and it corresponds to the cancellation of the spin connection by the background R-symmetry gauge field.
 
\section{The Cardy limit}
\label{sect:Cardy}

In this section, we analyze the Cardy limit $\tau \to \ii 0^+$ of the \emph{refined} topologically twisted index of $\cN=1$ theories that are associated with D3-branes at conical toric Calabi-Yau three-fold singularities.
When a four-dimensional $\cN=1$ theory is compactified on $S^2$ with a topological twist, it might flow to a family of $\cN = (0,2)$ SCFTs in the IR that are labeled by a set of magnetic fluxes $\fn_I$ parameterizing the twist.
The holographic dual to such a SCFT is a warped background AdS$_3\times_w Y_7$ in type IIB supergravity, where $Y_7$ is topologically a fibration of $Y_5$ over $S^2$, with $Y_5$ being the Sasaki-Einstein base of the Calabi-Yau cone. 

The right-moving central charge $c_r (\fn , r)$ and the gravitational anomaly $k = c_r - c_l$ of the $\cN = (0,2)$ SCFT read \cite{Benini:2012cz, Benini:2013cda}
\bea\label{chargesc}
 c_r (\fn , r) & = 3 \Tr \gamma_3 R^2(r_I) = - 3 \left[ \! \text{ dim}\, G + \sum_{I} \text{ dim}\,\fR_I ( \fn_I - 1 ) \left( r_I- 1 \right)^2 \right] \, , \\
 k (\fn) & = \Tr \gamma_3 = - \text{ dim}\, G - \sum_{I} \text{ dim}\,\fR_I ( \fn_I - 1 ) \, ,
\eea
where $R(r_I)$ is the matrix of R-charges for the fermionic fields in the  theory and $\gamma_3$ is the two-dimensional chirality operator.  The massless fermions arise as zero modes of four-dimensional fields.   The difference between the number of fermions of opposite chiralities is easily computed using the Riemann-Roch theorem and is equal to $-\text{ dim}\, G$ for the gaugino zero modes and  $ - \text{ dim}\,\fR_I ( \fn_I - 1 )$ for the zero modes of four-dimensional matter fields in a representation $\fR_I$  \cite{Benini:2012cz, Benini:2013cda}.\footnote{We  use the conventions of \cite{Benini:2012cz, Benini:2013cda} that differ by a sign in the definition of $\gamma_3$ compared to 
\cite{Hosseini:2016cyf}.}
The exact R-charges of the SCFT are obtained by extremizing $c_r (\fn , r)$ with respect to a generic assignment of trial R-charges $r_I$ satisfying
\be
 \sum_{I \in W_a} r_I=2 \, .
\ee
This principle is known as $c$-extremization and it has been successfully compared with the prediction of holography for a large class of twisted compactifications \cite{Benini:2012cz, Benini:2013cda, Benini:2015bwz}.\footnote{For further developments see \cite{Karndumri:2013iqa,Amariti:2016mnz,Hosseini:2016cyf,Amariti:2017cyd,Amariti:2017iuz,Hosseini:2018uzp,Couzens:2018wnk,Gauntlett:2018dpc,Hosseini:2019use}.}
In the large $N$ limit, for theories with a holographic dual, $k=0$ and $c_l=c_r$ \cite{Henningson:1998gx}.

In the following we will identify the modulus of the torus with the \emph{fictitious}%
\footnote{The elliptic genus is only counting extremal states and thus the temperature represented by $\im \tau$ is fictitious.}
inverse temperature $\beta$ and work in the Cardy limit
\be
 \beta \to 0 \quad \text{ with } \quad \beta \equiv - 2 \pi \ii \tau \, .
\ee
The refined topologically twisted index as $\beta \to 0$ can be written as
\bea
 \label{Cardy:RTTI}
 Z ( \fm , u ; \fn , \Delta , \omega | \beta) \sim \sum_{\fm \in \Gamma_\fh} \int_{\text{JK}} & \bigg( \prod_{\text{Cartan}} \ii \; \!\rd u \, \ex^{- \frac{\pi^2}{3 \beta}} \bigg)
 \prod_{\alpha \in G} \ex^{- \frac{1}{\beta} g_2 ( \alpha ( u ) ) - \frac{\omega^2}{8 \beta} \alpha ( \fm )^2} \\
 & \times \prod_{I} \prod_{\rho_I \in \fR_I} \ex^{\frac{1}{\beta} B_Ig_2 ( \rho_I (u) + \nu_I (\Delta) ) + \frac{\omega^2}{4 \pi^3 \beta} g_3 ( \pi ( B_I + 1 ) )} \, ,
\eea
where we used \eqref{eta:theta:near:q=1}, \eqref{1-loop:near:q=1}, and
\be
 \frac{\sign(B)}{\beta} \sum_{j = -\frac{|B| - 1}{2}}^{\frac{|B| - 1}{2}} g_2 ( \rho ( u ) + \nu_I (\Delta) +  j \omega ) = \frac{B}{\beta} g_2 (\rho(u) + \nu_I (\Delta) ) + \frac{\omega^2}{4 \pi^3 \beta} g_3 (\pi (B+1))
 \, .
\ee
The polynomial functions $g_s (u)$, $s=2,3,$ are defined in \eqref{g:functions}. We assumed that $\re ( \rho ( u ) + \nu_I (\Delta) +  j \omega)>0$. As we will see, this condition is satisfied 
on the relevant saddle point and in the regime of parameters that lead to a regular black hole, which in particular requires $\omega$ to be imaginary.

Let us now focus on the terms proportional to $\omega^2 / \beta$ in \eqref{Cardy:RTTI}, \ie\;
\be
 \label{omega:2:int}
 Z_\omega ( \fm , u ; \fn , \Delta | \beta ) = \prod_{\alpha \in G} \ex^{-\frac{\omega^2}{8 \beta} \alpha ( \fm )^2}
 \prod_{I} \prod_{\rho_I \in \fR_I} \ex^{\frac{\omega^2}{4 \pi^3 \beta} g_3 ( \pi ( B_I + 1 ) )} \, .
\ee
Remarkably, the dependence on gauge magnetic fluxes $\fm$ drops out of \eqref{omega:2:int} due to the following anomaly cancellation conditions:
\bea
 & \sum_{\alpha \in G} \alpha( \fm )^2 + \sum_I \sum_{\rho_I \in \fR_I} ( \nu_I(\fn) - 1 ) \rho_I(\fm)^2 = 0 \, , \quad && \U(1)_R\text{-gauge}^2 \, , \\
 & \sum_I \sum_{\rho_I \in \fR_I} \rho_I ( \fm )^3 = 0 \, , && \text{gauge}^3 \, , \\
 & \sum_I \sum_{\rho_I \in \fR_I} \rho_I ( \fm ) = 0 \, , && \text{gravitational}^2\text{-gauge} \, , \\
 & \sum_I \sum_{\rho_I \in \fR_I} \rho_I ( \fm ) \nu_I ( \fn )^2 = 0 \, , && \U(1)^2_R\text{-gauge} \, , \\
 & \sum_I \sum_{\rho_I \in \fR_I} \rho_I(\fm) \nu_I(\fn) = 0 \, , && \U(1)_R\text{-gravitational-gauge} \, ,
\eea
which are satisfied for all consistent theories of D3-branes at conical singularities.\footnote{Notice that, due to \eqref{Rfluxes}, the fluxes $\fn_a$ effectively parameterize an R-symmetry.}

We therefore find that
\be
 \label{Cardy:Z:omega}
 Z_\omega ( u ; \fn , \Delta | \beta ) = \prod_{I} \prod_{\rho_I \in \fR_I} \ex^{- \frac{\omega^2}{4 \pi^3 \beta} g_3 ( \pi \nu_I (\fn) )} \, .
\ee
Hence, the refined twisted index \eqref{Cardy:RTTI} can be further simplified to
\bea
 \label{Cardy:RTTI:anomaly:free}
 Z ( \fm , u ; \fn , \Delta , \omega | \beta) \sim Z_{0} ( \fm , u ; \fn , \Delta | \beta)
 \prod_{I} \prod_{\rho_I \in \fR_I} \ex^{- \frac{\omega^2}{4 \pi^3 \beta} g_3 ( \pi \nu_I (\fn) )} \, ,
\eea
where $Z_0$ is the \emph{unrefined} twisted index in the Cardy limit \cite{Hosseini:2016cyf}
\bea
 \label{Cardy:TTI}
 Z_0 ( \fm , u ; \fn , \Delta  | \beta) \sim \sum_{\fm \in \Gamma_\fh} \int_{\text{JK}} \bigg( \prod_{\text{Cartan}} \ii \; \!\rd u \, \ex^{- \frac{\pi^2}{3 \beta}} \bigg)
 \prod_{\alpha \in G} \ex^{- \frac{1}{\beta} g_2 ( \alpha ( u ) )}
 \prod_{I} \prod_{\rho_I \in \fR_I} \ex^{\frac{1}{\beta} B_I g_2 ( \rho_I (u) + \nu_I (\Delta) )} \, .
\eea
Notice that the correction term is independent of $u$, $\Delta$ and $\fm$, so we can easily write the partition function for generic $\omega$   if we know the  unrefined twisted index. 

The unrefined twisted index in the Cardy limit --- and at finite $N$ --- has been studied 
in details in \cite{Hosseini:2016cyf}. It can be evaluated by first resumming the geometric series associated with the gauge magnetic fluxes $\fm$.
Using the residue theorem, one can reduce the partition function to a sum over Bethe vacua, the critical points of the twisted superpotential $\cW(u;\Delta|\beta)$ obtained by dimensionally reducing the theory on 
$T^2$ \cite{Nekrasov:2014xaa,Benini:2015eyy,Closset:2017zgf}.\footnote{The Bethe vacua approach has been useful in the microscopic counting of states for many black holes and black strings in diverse dimensions \cite{Benini:2015eyy,Hosseini:2016cyf,Hosseini:2017fjo,Benini:2017oxt,Hosseini:2018uzp,Benini:2018ywd}.} In the Cardy limit, one particular vacuum dominates the partition function. For a theory of D3-branes at conical singularities with $\SU(N)$ gauge groups the distribution of eigenvalues was found in \cite{Hosseini:2016cyf}. It is the same for all gauge groups and reads  
\bea
 \label{BEA}
 u_i-u_j = - \frac{\ii \beta}{N} (i -j) \, \qquad\qquad i,j=1,\ldots, N \, .
\eea
As shown in \cite{Hong:2018viz}, this is actually a solution to the Bethe equations for arbitrary $\beta$.
The unrefined twisted index in the Cardy limit and at finite $N$  can then be compactly written as \cite{Hosseini:2016cyf}
\be
 \label{Cardy:formula}
 \log Z_0 ( \fn , \Delta | \beta) = \frac{\pi^2}{6 \beta} c_l \left ( \fn , \Delta/\pi \right) \, , \qquad \text{ as } \beta \to 0 \, ,
\ee
where $c_l ( \fn , \Delta/\pi )$   is the trial left-moving central charge of the $\cN=(0,2)$ SCFT in the IR.\footnote{It can be obtained from \eqref{chargesc} as  $c_l ( \fn , \Delta/\pi ) = c_r ( \fn , \Delta/\pi ) -k(\fn)$.} The result is consistent with the expectation that the topologically twisted index is computing the elliptic genus of the two-dimensional SCFT.  \eqref{Cardy:formula} is indeed nothing else than the supersymmetric version of {\it Cardy's formula} for the high-temperature behavior of the partition function of a CFT.
The result  \eqref{Cardy:formula} was obtained by using \eqref{eta:theta:near:q=1}, \eqref{1-loop:near:q=1} and assumes that
$\re (\Delta_I) \in [0,2\pi]$ and 
\be\label{constr0}
 \sum_{I \in W_a} \Delta_I = 2 \pi \, ,
\ee 
for each term $W_a$ in the superpotential.\footnote{The condition $\re (\Delta_I) \in [0,2\pi]$ implies that $\sum_{I \in W_a} \Delta_I$ cannot be zero. Among the possible choices allowed by  \eqref{constrDelta}, only $\sum_{I \in W_a} \Delta_I = 2\pi$ leads to a physically acceptable solution \cite{Hosseini:2016cyf}.}
We see that $\Delta_I/\pi$ effectively parameterize a choice of R-charges for the fields of the theory.
Notice also that our working condition  $\re ( \rho ( u ) + \nu_I (\Delta) +  j \omega)>0$ is satisfied if $\omega$ is imaginary, since
the distribution of eigenvalues satisfies \eqref{BEA}. 

Plugging back \eqref{Cardy:formula} into \eqref{Cardy:RTTI:anomaly:free} we finally arrive at the following expression for the Cardy limit of the refined twisted index
\bea
 \label{Cardy:formula:RTTI}
 \log Z ( \fn , \Delta , \omega  | \beta ) & = \frac{\pi^2}{6 \beta} c_l ( \fn , \Delta/\pi )
 - \frac{\omega^2}{4 \pi^3 \beta} \sum_{I} \text{ dim}\,\fR_I \, g_3 ( \pi \fn_I ) \\
 & = \frac{\pi^2}{6 \beta} c_l ( \fn , \Delta/\pi )
 - \frac{\omega^2}{24 \beta} \left( \Tr R^3 (\fn) - \Tr R (\fn) \right) \\
 & = \frac{\pi^2}{6 \beta} c_l ( \fn , \Delta/\pi )
 - \frac{( 2 \omega)^2}{27 \beta} \left( 3 c(\fn) - 2 a(\fn) \right) \, , \qquad \text{ as } \beta \to 0 \, .
\eea
In order to write \eqref{Cardy:formula:RTTI} we introduced the traces over four-dimensional fermionic fields
\bea
 \Tr R^3(r_I)  &=   \text{ dim}\, G + \sum_{I} \text{ dim}\,\fR_I  ( r_I  - 1 )^3 \, , \\
 \Tr R(r_I) &=   \text{ dim}\, G + \sum_{I} \text{ dim}\,\fR_I  ( r_I  - 1 ) \, ,
\eea
where $r_I$ is the R-charge of the $I$-th chiral multiplet, related by
\bea
 \label{a:c:trial}
  a(r_I) = \frac{9}{32} \Tr R^3(r_I) - \frac{3}{32} \Tr R(r_I) \, , \qquad c(r_I) = \frac{9}{32} \Tr R^3(r_I) - \frac{5}{32} \Tr R(r_I) \, ,
\eea
to the trial central charges of the four-dimensional $\cN=1$ theory \cite{Anselmi:1997am}. Notice that, due to the twisting condition \eqref{Rfluxes}, the fluxes $\fn_I$ parameterize an integer choice of R-charges for the chiral fields of the theory.   

In the large $N$ limit, $a=c$ and $c_l=c_r$.  We can thus further simplify \eqref{Cardy:formula:RTTI} and write
\be
 \label{Cardy:largeN:cr:a}
 \log Z (\fn, \Delta , \omega | \beta) = \frac{\pi^2}{6 \beta} \bigg( c_r (  \fn , \Delta/\pi ) - \frac{8 \omega^2}{9 \pi^2} a( \fn ) \bigg) \, , \qquad \text{ as } \beta \to 0\, ,\, N \gg 1 \, .
\ee
For theories associated with D3-branes at  toric Calabi-Yau singularities, we can always choose a convenient parameterization of the general R-charges of the chiral fields
satisfying \eqref{Rfluxes} in terms of a minimal set of $d$ quantities $r_a$ with
\be
 \label{toric:R-charge}
 \sum_{a=1}^d r_a=2 \, ,
\ee
where $d$ is the number of global symmetries of the theory \cite{Butti:2005vn,Benvenuti:2005ja}.
In this convenient parameterization, the central charge is a homogeneous function of degree three \cite{Benvenuti:2006xg}
\be
 a (r) \equiv \frac{9 N^2}{64} \sum_{a,b,c = 1}^{d} c_{abc} r_a r_b r_c \, ,
\ee
where $c_{abc}$ are proportional to the 't Hooft anomaly coefficients. For a toric quiver $c_{abc}=|\det (v_a,v_b,v_c)|$, where $v_a\in \mathbb{Z}^3$ are the integer vectors defining the toric diagram.%
\footnote{We redefined $c_{abc}^{\text{here}} \equiv \frac{2}{N^2} c_{a b c}^{\text{there}}$ in comparison with \cite[Eq.\,(2.7)]{Hosseini:2019use}.}
Moreover, as shown in \cite{Hosseini:2016cyf,Hosseini:2019use},  in the large $N$ limit the two-dimensional central charge can be written as
\be
 \label{2d}
 c_r(\fn,r) = -\frac{32}{9} \sum_{a=1}^d \fn_a \frac{\partial a(r_a)}{\partial r_a} = - \frac{3 N^2}{2} \sum_{a,b,c=1}^d c_{a b c} \fn_a r_b r_c \, .
\ee

It is convenient, also for comparison with other microstate counting for black holes and black strings, to  express all quantities in terms of
the \emph{on-shell} value of the effective twisted superpotential of the theory, $\wt\cW (\Delta | \beta)$,
which is by definition $\cW(u ; \Delta | \beta)$ evaluated on the Bethe vacuum solution \eqref{BEA}.
It is related to the $a$ central charge of the four-dimensional $\cN=1$ theory as \cite{Hosseini:2016cyf}
\be
 \label{W:a}
 \wt \cW (\Delta | \beta ) = \frac{16 \pi^3 \ii}{27 \beta} a ( \Delta/\pi )
 = \frac{\ii N^2}{12 \beta} \sum_{a,b,c} c_{abc} \Delta_a \Delta_b \Delta_c \, ,
 \qquad \text{ for } N \gg 1\, .
\ee
Combining \eqref{Cardy:largeN:cr:a} and \eqref{2d} we can finally write the compact expression
\be
 \label{logZ:W}
 \log Z (\fn, \Delta , \omega | \beta) = \ii \sum_{a=1}^{d} \fn_a \frac{\partial \wt \cW (\Delta | \beta)}{\partial \Delta_a}
 + \frac{\ii \omega^2}{24} \sum_{a , b , c=1}^{d} \fn_a \fn_b \fn_c \frac{\partial^3 \wt \cW (\Delta | \beta)}{\partial \Delta_a \partial \Delta_b \partial \Delta_c} \, .
\ee
The first term in \eqref{logZ:W} was already derived in \cite[sect.\,5.3]{Hosseini:2016cyf}.

We now present two simple examples, $\cN=4$ super Yang-Mills and the Klebanov-Witten theory \cite{Klebanov:1998hh}. 

\paragraph*{$\cN = 4$ super Yang-Mills.}

Our first example is the $\cN=4$ super Yang-Mills (SYM) theory whose $\cN=1$ quiver is depicted below.
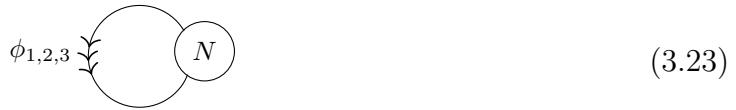
\be
 \begin{aligned}
 \label{SYM:quiver}
 \begin{tikzpicture}[font=\footnotesize, scale=0.9]
 \begin{scope}[auto,%
  every node/.style={draw, minimum size=0.5cm}, node distance=2cm];
 \node[circle]  (UN)  at (0.3,1.7) {$N$};
 \end{scope}
 \draw[decoration={markings, mark=at position 0.45 with {\arrow[scale=1.5]{>}}, mark=at position 0.5 with {\arrow[scale=1.5]{>}}, mark=at position 0.55 with {\arrow[scale=1.5]{>}}}, postaction={decorate}, shorten >=0.7pt] (-0,2) arc (30:341:0.75cm);
 \node at (-2.1,1.7) {$\phi_{1,2,3}$};
 \end{tikzpicture}
 \end{aligned}
\ee
Here the circular node is a $\SU(N)$ gauge group and the loop around it denotes the adjoint chiral multiplets $\phi_a$, $a=1,2,3,$ interacting through the cubic superpotential
\be
 \label{SYM:W}
 W = \Tr \left( \phi_3 \left[ \phi_1, \phi_2 \right] \right) \, .
\ee
We denote the chemical potentials associated to the chiral fields $\phi_a$ by $\Delta_a$, and assign the flux $\fn_a$ to each chiral field.
Then, the invariance of \eqref{SYM:W} under the global symmetries of the theory imposes the constraint
\be
 \sum_{a = 1}^{3} \Delta_a = 2 \pi \, .
\ee
A similar constraint exists on the magnetic fluxes $\fn_a$, see \eqref{Rfluxes}.
The trial central charges, at finite $N$, are easily computed from \eqref{a:c:trial} and read%
\footnote{For $\cN=4$ SYM we have $\Tr R (r_a) = 0$.}
\be
 \label{SYM:a(n)}
 a (r) = c(r) = \frac{27 (N^2-1)}{32} r_1 r_2 r_3 \, ,
\ee
with $\sum_{a=1}^3 r_a = 2$. The only nonzero anomaly coefficients are $c_{123}=1$ and permutations thereof.
For $\cN=4$ SYM compactified on $S^2$ the gravitational anomaly of the two-dimensional $\cN = (0,2)$ theory is zero, $k=0$, and thus $c_l = c_r$.
The trial right-moving central charge $c_r$ is given by \eqref{chargesc},
\be
 \label{SYM:c2d:trial}
 c_{r}  ( \fn , r ) = - 3 ( N^2 - 1)  ( r_1 r_2 \fn_3 + r_2 r_3 \fn_1 + r_1 r_3 \fn_2 ) \, .
\ee
Extremizing \eqref{SYM:c2d:trial} with respect to $r_{a}$, $a=1,2,$ we obtain
\be
 \label{SYM:bar:Delta}
 \bar r_{a}^{\?\cN=4} = \frac{4 \fn_a (\fn_a - 1)}{\Theta_{\cN=4}} \, , \qquad a=1,2 \, ,
\ee
where we defined
\be
 \label{SYM:theta}
 \Theta_{\cN=4} \equiv \fn_1^2 + \fn_2^2 + \fn_3^2 - 2 (\fn_1 \fn_2 + \fn_1 \fn_3 + \fn_2 \fn_3) \, .
\ee
Plugging \eqref{SYM:bar:Delta} back into \eqref{SYM:c2d:trial} we find the exact central charge of the 2D CFT as
\be
 \label{SYM:cr}
 c_r^{\text{CFT}}(\fn) = 12 (N^2 - 1) \frac{\fn_1 \fn_2 \fn_3 }{\Theta_{\cN=4}} \, .
\ee
Finally, the refined twisted index in the Cardy limit is given by \eqref{Cardy:formula:RTTI},
\be
 \label{SYM:RTTI}
 \log Z_{\cN=4} (\fn , \Delta | \beta) = - \frac{(N^2 - 1)}{8 \beta} \left( 4 \Delta_1 \Delta_2 \fn_3 + 4 \Delta_2 \Delta_3 \fn_1 + 4 \Delta_1 \Delta_3 \fn_2 + \omega^2 \fn_1 \fn_2 \fn_3 \right) .
\ee

\paragraph*{The Klebanov-Witten theory.}

Our second example is the twisted compactification of the Klebanov-Witten theory \cite{Klebanov:1998hh}.
It is an $\cN=1$ gauge theory with gauge group $\SU(N) \times \SU(N)$ and two pairs of bi-fundamental chiral fields $\{A_i, B_j\}$, $i , j = 1,2,$ in the representations $({\bf N},\overline{{\bf N}})$ and $(\overline{{\bf N}},{\bf N})$, respectively.
The matter content is described by the quiver diagram
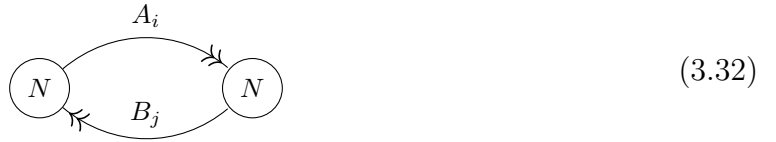
\be
 \begin{aligned}
 \label{KW:quiver}
 \begin{tikzpicture}[baseline, font=\footnotesize, scale=0.8]
 \begin{scope}[auto,%
  every node/.style={draw, minimum size=0.5cm}, node distance=2cm];
 \node[circle] (USp2k) at (-0.1, 0) {$N$};
 \node[circle, right=of USp2k] (BN)  {$N$};
 \end{scope}
 \draw[decoration={markings, mark=at position 0.9 with {\arrow[scale=1.5]{>}}, mark=at position 0.95 with {\arrow[scale=1.5]{>}}}, postaction={decorate}, shorten >=0.7pt]  (USp2k) to[bend left=40] node[midway,above] {$A_{i}$} node[midway,above] {} (BN) ; 
 \draw[decoration={markings, mark=at position 0.1 with {\arrow[scale=1.5]{<}}, mark=at position 0.15 with {\arrow[scale=1.5]{<}}}, postaction={decorate}, shorten >=0.7pt]  (USp2k) to[bend right=40] node[midway,above] {$B_{j}$}node[midway,above] {}  (BN) ;  
 \end{tikzpicture}
 \end{aligned}
\ee
and there is a quartic superpotential
\be
 \label{KW:W}
 W = \Tr ( A_1 B_1 A_2 B_2 - A_1 B_2 A_2 B_1 ) \, .
\ee
We assign chemical potentials $\Delta_a$ and fluxes $\fn_a$, $a=1,\ldots,4,$ to the fields $\{A_1,A_2,B_1,B_2\}$ in this order.
The invariance of the superpotential \eqref{KW:W} under the global symmetries of the theory imposes the constraint
\be
 \sum_{a = 1}^{4} \Delta_a = 2 \pi \, .
\ee
The twisting condition \eqref{Rfluxes} also reads
\be
 \sum_{a = 1}^{4} \fn_a = 2 \, .
\ee
The trial central charges of the four-dimensional theory (at finite $N$) are given by \eqref{a:c:trial},
\bea
 \label{KW:a(n)}
 a (r) = \frac{27 N^2}{32} \sum_{a < b < c} r_a r_b r_c - \frac{3}{8} \, , \qquad
 c (r) = \frac{27 N^2}{32} \sum_{a < b < c} r_a r_b r_c - \frac{1}{4} \, ,
\eea
where again $r_a$ parameterize an R-symmetry of the theory and $\sum_{a=1}^4 r_a = 2$.
For later convenience we introduce the function
\be
 \Theta_{\text{KW}} ( \fn ) = \frac{\varsigma( \fn )}{\vartheta( \fn )} \, ,
\ee
where we defined
\be
 \varsigma ( \fn ) \equiv {\frac{1}{3} \left( 4 \sum_{a = 1}^4 \fn_a^3-\sum _{a , b , c = 1}^4 \fn_a \fn_b \fn_c \right) \sum_{a<b<c} \fn_a \fn_b \fn_c} \, ,
 \qquad \vartheta ( \fn )  \equiv \sum_{ \substack{a<b \\ (\neq c)}} \fn_a \fn_b \fn_c^2 \, .
\ee
The trial central charges of the two-dimensional $\cN = (0,2)$ theory in the IR read \eqref{chargesc},
\bea
 \label{KW:2d}
 & c_r (\fn , r) = - 3 N^2 \sum_{\substack{a<b \\ (\neq c)}} r_a r_b \fn_c + 6 \, , \qquad k = 2 \, , \\
 & c_l (\fn , r) = c_r (\fn , r) - k \, .
\eea
Extremizing $c_r$ in \eqref{KW:2d} with respect to $r_a$, $a=1,2,3$, under the constraint $\sum_{a=1}^4 r_a = 2$, yields the critical points
\be
 \bar r_a^{\?\text{KW}} = \frac{12}{\fn_a} \frac{\fn_a^2 \left( \fn_a^2 - \frac12 \sum\limits_{a = 1}^4 \fn_a^2 \right) - \sum\limits_{a<b<c<d} \fn_a \fn_b \fn_c \fn_d}
 {4 \sum\limits_{a = 1}^4 \fn_a^3 - \sum\limits_{a , b , c = 1}^4 \fn_a \fn_b \fn_c} \, , \qquad a=1,2,3 \, .
\ee
Substituting this back into \eqref{KW:2d} we obtain
\be
 c_r^{\text{CFT}} (\fn) = \frac{12 N^2}{\Theta_{\text{KW}}} \sum_{a < b < c} \fn_a \fn_b \fn_c + 6 \, .
\ee
Finally, the refined twisted index in the Cardy limit is given by \eqref{Cardy:formula:RTTI},
\be
 \label{KW:RTTI}
 \log Z_{\text{KW}} (\fn , \Delta | \beta) = - \frac{N^2}{2 \beta} \sum_{\substack{a<b \\ (\neq c)}} \Delta_a \Delta_b \fn_c
 + \frac{2 \pi^2}{3 \beta}
 - \frac{\omega^2 N^2}{8\beta } \sum_{a < b < c} \fn_a \fn_b \fn_c \, .
\ee

\section{Rotating black strings in AdS$_5$}
\label{sect:rotating:strings}

In this section we consider rotating black string solutions in five-dimensional gauged supergravity.\footnote{To be more precise, we consider black strings with spherical topology and rotation along the sphere. For rotating {\it hyperbolic} black strings with noncompact horizon, see \cite{Azzola:2018sld}.} Some of these solutions can be  embedded in AdS$_5\times S^5$ and represent the holographic dual of the field theory setting discussed in the previous section for $\cN=4$ SYM. It is  convenient, both from supergravity and field theory perspective, to perform a dimensional reduction to a four-dimensional black hole which carries  momentum along the compactification circle. As we will see, it is the entropy of such black holes that  can be directly compared with the field theory microscopic counting performed in section \ref{sect:Cardy}.

As shown in \cite{Hristov:2014eza}, asymptotically AdS$_5$ black string solutions of five-dimensional gauged supergravity can be reduced to four-dimensional solutions in a gauged supergravity \emph{without} a maximally symmetric vacuum solution.
In four dimensions the resulting black hole solutions can be thought of as having runaway asymptotics of the hvLif type with particular exponents as described in more detail in \cite[sect.\,3.1]{Hristov:2018spe}. A similar construction was successfully used to relate the Gutowski-Reall black holes in AdS$_5$ and their generalizations to four-dimensional black holes \cite{Hosseini:2017mds}.

Here we want to  generalize the original construction presented in \cite{Hristov:2014eza} by allowing for rotation. Since the 5D/4D relation actually provides a one-to-one map, we take the approach of first writing down the solutions in four dimensions, and then uplifting them to five dimensions. The four-dimensional  black holes preserve supersymmetry with a twist, and the general solutions of this type that include rotation were recently found in \cite{Hristov:2018spe} for arbitrary symmetric models of gauged supergravity with vector multiplets. Here we will  discuss a particular subclass of symmetric models and will explicitly write down the relevant solutions of \cite{Hristov:2018spe}. We will discuss in particular the STU  model that can be embedded in AdS$_5\times S^5$. Unfortunately, more general compactifications AdS$_5\times {\rm SE}_5$ based on Sasaki-Einstein manifolds are not included in this construction since the known consistent truncations contain hypermultiplets, see \cite{Gauntlett:2007ma,Cassani:2010na,Bena:2010pr}. Additionally, these truncations do not yet include all abelian isometries of the internal spaces to allow for an interesting comparison with field theory.  

\subsection{The 5D/4D relation}
Consider a  five-dimensional $\cN=2$ gauged supergravity theory coupled to $n_V$ vector multiplets. The bosonic fields are the metric, $n_{\rm V}$ abelian gauge fields $A_{(5)}^i$
and real scalar fields $L^i$ $(i = 1 , \ldots , n_{\rm V})$ parameterizing the manifold $\frac 1 6 c_{ijk} L^i L^j L^k=1$. Here $c_{ijk}$ is a fully symmetric tensor appearing in the Chen-Simons terms and it corresponds to the 't Hooft anomaly coefficients of the dual $\cN=1$ four-dimensional CFT \cite{Ferrara:1998ur,Benvenuti:2006xg}.  The   rules for reducing the bosonic fields along the circle $x^5$ are the following \cite{Andrianopoli:2004im,Gaiotto:2005gf,Behrndt:2005he,Cardoso:2007rg}:%
\footnote{In our conventions the five-dimensional and four-dimensional gauge fields are related by $A^i_{(5)} = \sqrt{2} A^i_{(4)}$ and the gauge coupling constants in the respective gauged supergravity by $g_{(5)} = \frac{1}{\sqrt{2}} g_{(4)}$.}
\bea
 \label{5D:4D:relation}
 \rd s_{(5)}^2 & = \ex^{2 \varphi} \, \rd s_{(4)}^2 + \ex^{- 4 \varphi} \, \left( \rd x^5 - \sqrt{2} A_{(4)}^0 \right)^2 \, , \qquad && \rd x^5 = \rd y \, , \\
 A_{(5)}^i & = \sqrt{2} A_{(4)}^i + \re z^i \left( \rd x^5 - \sqrt{2} A_{(4)}^0 \right) \, , \\
 L^i & = \ex^{2 \varphi} \im z^i \, , && \ex^{-6 \varphi} = \frac16 c_{i j k}  \im z^i  \im z^j  \im z^k \, .
\eea
Here $\rd s_{(4)}^2$ denotes the four-dimensional line element,
$A_{(4)}^I$ $(I = 0, i )$ are the four-dimensional abelian gauge fields and $z^i$
are the complex scalar fields in four dimensions. The four-dimensional theory has $n_{\rm V}$ abelian vector multiplets,
parameterizing a special K\"ahler manifold $\cM$ with metric $g_{i \bar{j}}$,
in addition to the gravity multiplet (thus a total of $n_{\rm V}+1$ gauge fields and $n_{\rm V}$ complex scalars).
The scalar manifold is defined by the prepotential $\cF \left( X^I \right)$,
which is a homogeneous holomorphic function of sections $X^I$,
\bea
 \label{eq:prepotential}
 \cF \left( X^I \right) = \frac16 \frac{c_{i j k} X^i X^j X^k}{X^0} \, .
\eea 
In $\cN = 2$ gauged supergravity in four dimensions the $\U(1)_R$ symmetry, rotating the gravitini,
is gauged by a linear combination of the abelian gauge fields.
The coefficients are called Fayet-Iliopoulos (FI) parameters $g_I$
and $n_{\text{V}}$ of them can be directly read off from the five-dimensional theory: $g_i$.%
\footnote{In consistent models one can always apply an electric-magnetic duality transformation so that the corresponding gauging becomes purely electric, \ie, $g^I = 0$.}
The last coefficient, $g_0$, measuring how the Kaluza-Klein gauge potential $A^0_{(4)}$ enters the R-symmetry,
we choose to be vanishing since a standard Kaluza-Klein reduction suffices our purposes here (as opposed to a more general Scherk-Schwarz reduction, see \cite{Hosseini:2017mds}).

\subsection{The 4D rotating solutions: asymptotics and near horizon}

The definition of the four-dimensional models, as well as the subsequent solution, are most commonly written in a duality covariant language in terms of symplectic vectors. We also extensively use the formalism where quantities are written in terms of the quartic invariant $I_4$ as explained in \cite{Hristov:2018spe}. In particular, the model we are interested in is specified by the prepotential \eqref{eq:prepotential}, where the constant tensor $c_{i j k}$ further satisfies the identity
\begin{equation}\label{eq:symm-cub}
 \frac43\?\delta_{i(l}\? c_{mpq)} = c_{ijk}\?c_{j^\prime (l m} \?c_{pq)k^\prime}\?\delta^{j j^\prime}\?\delta^{k k^\prime} \, ,
\end{equation}
from the requirement that the scalar manifold is a symmetric space. This guarantees the existence of a rank 4 symplectic tensor, which upon contraction with a generic charge vector $\Gamma = \{p^I ; q_I\}$ defines the quartic form 
\begin{eqnarray}\label{I4-ch-0}
	I_4(\Gamma) = - (p^0 q_0 + p^i q_i)^2 + \frac{2}{3} q_0 c_{ijk} p^i p^j p^k - \frac{2}{3} p^0 c^{ijk} q_i q_j q_k + c_{ijk} p^j p^k c^{ilm}q_l q_m \, , 
\end{eqnarray}
invariant under symplectic transformations. Subsequently we will use the quartic invariant form coming from the contraction of the different symplectic objects in order to fully characterize the solution. In the above formula, indices are raised and lowered with the K\"{a}hler metric on the scalar manifold that depends on the prepotential \eqref{eq:prepotential}, see \cite{Andrianopoli:1996cm} for the complete set of conventions we follow. The condition \eqref{eq:symm-cub} ensures that the inverse tensor $c^{ijk}$ also has constant entries and therefore the quartic form is not scalar dependent.\footnote{It is not difficult to see that the anomaly coefficients of a theory of D3-branes at a toric singularity satisfy \eqref{eq:symm-cub} only in the case of $\cN=4$ SYM. This is another reason why the class of solutions discussed in this section cannot be immediately used for studying black strings in general Sasaki-Einstein compactifications.}

The  FI parameters  can be encoded in a gauging vector
\begin{equation}\label{eq:gaugingvector}
G = \{g^I= 0; g_0 = 0, g_i \}\, ,
\end{equation}
and we can evaluate the quartic invariant and its first derivative on the vector $G$,
\be
I_4(G) = 0,\quad (I_4')_0 (G) = - \frac{\partial I_4 (G)}{\partial g^0} = \frac23\, c^{i j k} g_i g_j g_k\,,
\ee
where the first derivative $I'_4$ also transforms covariantly as a vector. 
The vanishing of $I_4(G)$ means that no AdS$_4$ vacuum can exist in this supergravity model, and instead the nonzero $I_4' (G)$ signifies the hvLif asymptotics already anticipated to uplift to AdS$_5$ when using the rules \eqref{5D:4D:relation}. 

Now, let us focus our attention to the near horizon geometry of the four-dimensional black hole we are interested in.
We want to generalize the solution discussed in \cite{Hristov:2014eza} and first found in \cite{Cacciatori:2009iz} (and in \cite{Hristov:2014hza} for the case with extra electric charges) to include angular momentum $\cJ$.
Therefore, we are interested in a slightly restricted set of electromagnetic charges,
\begin{equation}
 \label{eq:chargevector}
 \Gamma = \{ p^0 = 0,p^i; q_0, q_i \}\, ,
\end{equation}
which, together with $\cJ$ will eventually specify the complete supersymmetric solution. In terms of the five-dimensional rotating string, the $p^i$ and $q_i$ are the magnetic and electric charges, while $q_0$ is the momentum added along the compactification direction.\footnote{Comparing with  the dual field theory setting of  section \ref{sect:Cardy}, $p^i$ are associated with the fluxes $\fn_a$, the electric charges $q_i$ are conjugated to the chemical potentials $\Delta_a$ and $q_0$ is conjugated to $\beta$. The precise dictionary is spelled out   in section \ref{sect:holography}.}
The general near horizon geometry with a twist and rotation was written down in \cite[sect.\,4]{Hristov:2018spe}.
Here we just repeat the main ingredients and solve the attractor equation that determines explicitly all quantities. Note that the chosen set of charges in general will lead to a non-trivial NUT charge in the solution, so below we will restrict one of the electric charges $q_i$ in a specific way in order to ensure regularity and write down the full solution in a compact form. 

The metric can be written in the form
\begin{equation}\label{eq:metr-bps-atrr}
\rd s^2_4 = -\ex^{2\?\u} \left(r \? \rd t +  \omega_0 \right)^2 + \ex^{-2\?\u}\?\left( \frac{\rd r^2}{r^2} + {\rm v}^2 \left( \frac{\rd \theta^2}{\Delta(\theta)} + \Delta(\theta) \sin^2(\theta)\, \rd \phi^2 \right) \! \right) ,
\end{equation}
where
\begin{equation}
 \ex^{-2\?\u} = \sqrt{I_4({\cal I}_0)} \, , \quad {\rm v}\?{\cal I}_0 = {\cal H}_0 + {\rm j}\? G\? \cos (\theta) \,,\quad {\rm v} =  \Iprod{G}{{\cal H}_0} = g_i \beta^i\, .
\end{equation}
In the last equality we already used the definition of symplectic inner product and the parameterization of the vector ${\cal H}_0$,
\be
	{\cal H}_0 = \{ 0,\beta^i;\beta_0, \beta_i \}\, ,
\ee
parallel to the charge vector $\Gamma$.
All remaining metric functions, as well as the scalars, are uniquely fixed in terms of the vector ${\cal H}_0$. For the choice of gauging and electromagnetic charges, we find from \cite{Hristov:2018spe} 
\be
	\Delta(\theta) = 1\, , \qquad \omega_0 = - \frac{{\rm j}}{\rm v} \sin^2 (\theta) \? {\rm d} \phi \, ,
\ee
where we already imposed regularity of the metric, \ie\;vanishing NUT charge, which we check below, and used
\be
	 I_4 (\Gamma, G, G, G) = I_4 ({\cal H}_0, G, G, G) = 0 \ .
\ee
The symplectic sections at the horizon, in a suitable gauge, are fixed by
\be
	\label{eq:hor-sec}
	\{X^I; F_I \} = - \frac{1}{2 \sqrt{I_4({\cal I}_0)}}\? I_4'({\cal I}_0) + \ii\? {\cal I}_0\, ,
\ee
and we can recover the physical scalars $z^i = X^i/X^0$.
We need to impose one standard constraint among the magnetic charges stemming from the twisting condition,
\be
 \label{eq:final-sph}
\Iprod{G}{\Gamma} = g_i p^i = -1\, .
\ee
Finally, the main attractor equation to be solved for ${\cal H}_0$ in terms of the charges reads
\be
\label{eq:attr-fin}
  \Gamma = \frac{1}{4}\? I^\prime_4\left({\cal H}_0, {\cal H}_0, G \right) + \frac{1}{2}\? {\rm j}^2\? I^\prime_4\left( G \right)\,,
\ee
together with the equation determining the angular momentum
\begin{equation}
 \label{eq:J-general-0}
 {\cal J} = - \frac{{\rm j}}{2}\?\Big( \Iprod{I_4^\prime(G)}{I_4^\prime(\cH_0)} - \frac12\? I_4(\cH_0, \cH_0, G, G)\?\Iprod{G}{\cH_0} \rule[.1cm]{0pt}{\baselineskip} \Big) \, .
\end{equation}
In principle, \eqref{eq:attr-fin} and \eqref{eq:J-general-0} can be solved for $\cH_0$ and ${\rm j}$ in terms of the conserved charges $\Gamma, {\cal J}$. This determines the complete solution.
However, the system of equations is not always easy to simplify in full generality.
In order to do so here, we impose the following restriction on the charges,
\be
	\label{eq:I4:id:zero}
	I_4(\Gamma, \Gamma, \Gamma, G) = 0\, ,
\ee
which also ensures that we have a vanishing NUT charge. The above relation will generally fix one of the electric charges $q_i$ in terms of other conserved charges.
Using \eqref{eq:I4:id:zero}, together with $I_4 (G) = 0$ and a multitude of special identities of the quartic invariant, spelled out in \cite[app.\,A.3]{Halmagyi:2014qza},
we can find a general solution to \eqref{eq:attr-fin} and then plug it in \eqref{eq:J-general-0} to finally arrive at the solution
\bea
	\cH_0  & = - \frac{1}{2 \sqrt{\Pi}}\? \left(\frac12\? I_4'(\Gamma, \Gamma, G) + \frac{(I_4(\Gamma) + {\cal J}^2)}{\Theta}\? I_4'(G)\right) \, , \\
	{\rm j} & = - \frac{\cal J}{\sqrt{\Pi}} \, , \qquad {\rm v} =  \Iprod{G}{{\cal H}_0} = \frac{\Theta}{\sqrt{\Pi}} \, ,
\eea
where we defined for shorthand the combinations
\be
	\Theta \equiv  -\frac14\? I_4(\Gamma, \Gamma, G, G) \, , \qquad \Pi \equiv  \frac12\? \Iprod{I_4'(\Gamma)}{I_4'(G)} - \frac14\? I_4(\Gamma, \Gamma, G, G) \, .
\ee
Note that now we can evaluate any physical quantity of the solution in terms of the conserved charges.
The near horizon solution can be uniquely extended to fix the full black hole flow to the asymptotic region using the formulae given in \cite{Hristov:2018spe}.
Here we are particularly interested in the entropy, which in terms of the parameters $\cH_0$ and ${\rm j}$ has the simple expression
\be
	S = \frac{A}{4\? G_{\text{N}}} = \frac{\pi}{G_{\text{N}}} \sqrt{I_4(\cH_0) - {\rm j}^2}\, ,
\ee
and in terms of the conserved charges, using more $I_4$ identities from \cite{Halmagyi:2014qza}, 
\be
	S = \frac{\pi}{G_{\text{N}}}\? \sqrt{\frac{- I_4(\Gamma) - {\cal J}^2}{\Theta}}\, .
\ee
We can also define the chemical potential conjugate to the angular momentum as in \cite{Hristov:2018spe},
\be
	w \equiv \frac{{\rm j}}{{\rm v} \sqrt{I_4(\cH_0) - {\rm j}^2}} = - \frac{{\cal J}}{\sqrt{\Theta (- I_4(\Gamma) - {\cal J}^2)}} \, .
\ee
We finish the discussion of the general near horizon solution with the formula determining the electric gauge fields,
\be
	A^I_{(4)} = - \frac{\Theta^5\? I_4'({\cal I}_0)^I}{2 \Pi^2\? \Xi (\theta)} (r\? {\rm d} t + \omega_0) - p^I \cos (\theta) {\rm d} \phi \, ,
\ee
where we have used the solution for ${\cal H}_0$ and ${\rm j}$ to simplify the formula, and for brevity we defined the quantity
\be
	\Xi (\theta) \equiv (- I_4(\Gamma) - {\cal J}^2) + \frac{\Theta}{\Pi} {\cal J}^2 \sin^2 (\theta) \, .
\ee

\subsection{STU model}
\label{sect:STU:model}
In the special case when $n_{\text{V}} = 3$ with $c_{1 2 3} = 1$ (and cyclic permutations), and gaugings $g_i =1$, $i=1,2,3$, we have the so-called STU model.
Upon uplift to five dimensions, it is also embeddable in maximal gauged supergravity
and follows from reduction of type IIB supergravity on S$^5$. 

In this model the rotating solutions we described above are a generalization of the black string solutions of \cite{Benini:2013cda}. In this case we find that
\bea
\label{I4-stu}
I_4^{\text{STU}}(\Gamma) &=  4 \,q_0 p^1 p^2 p^3 - \sum_{i = 1}^3 (p^i q_i)^2  +2 \sum_{i<j} q_i p^i q_j p^j \, , \\
\Theta^{\text{STU}} &=  (p^1)^2 + (p^2)^2 + (p^3)^2 - 2(p^1 p^2 + p^1 p^3 + p^2 p^3)\, , \\
\Pi^{\text{STU}} &= (-p^1+p^2+p^3) (p^1-p^2+p^3) (p^1+p^2-p^3)\, ,
\eea
under the constraint
\be\label{eq:STUmagneticconstraint}
p^1 + p^2 + p^3 = -1\, ,
\ee
in exact accordance with \cite{Benini:2013cda} for the case of spherical topology, $\fg=0$. Notice that
\be
 \Theta^{\text{STU}} - \Pi^{\text{STU}} = 8\? p^1 p^2 p^3 \, .
\ee

\subsubsection{Magnetic case}
Let us first for simplicity consider the magnetic case, $q_i = 0$. The constraint \eqref{eq:I4:id:zero} is then satisfied automatically.
The Bekenstein-Hawking entropy in this case reads
\be
	\label{STU:entropy}
	S^{\text{STU}} = \frac{\pi}{G_{\text{N}}}\? \sqrt{\frac{-4\? q_0 p^1 p^2 p^3- {\cal J}^2}{\Theta^{\text{STU}} } }\, ,
\ee
and the region of positivity (the charge parameter space where regular solutions exist) was analyzed and shown to be non-empty in \cite{Benini:2013cda}. The inclusion of the angular momentum naturally decreases or increases this region dependeing on the sign of $\Theta^{\text{STU}}$. Note however that due to the constraint \eqref{eq:STUmagneticconstraint} the solution does not allow a limit of vanishing charges.

If we explicitly evaluate the sections on the horizon from \eqref{eq:hor-sec} as functions of the spherical coordinate $\theta$, we obtain
\bea
& X^0 (\theta) = \frac{2\? p^1 p^2 p^3}{\sqrt{\Theta^{\text{STU}}\? \Xi^{\text{STU}}(\theta)}} \, , \\
& X^i (\theta) =  \frac{ p^i {\cal J} \cos (\theta)}{\sqrt{\Theta^{\text{STU}}\? \Xi^{\text{STU}}(\theta)}} +\ii\? \frac{p^i (1+2 p^i )}{\Theta^{\text{STU}}} \, , \qquad i = 1 , 2 , 3 \, ,
\eea
where in the absense of electric charges $q_i$ we find
\be
	\Xi^{\text{STU}} (\theta) \equiv (-4\? q_0 p^1 p^2 p^3 - {\cal J}^2) + \frac{\Theta^{\text{STU}}}{\Pi^{\text{STU}}} {\cal J}^2 \sin^2 (\theta)\, .
\ee
Moreover, the chemical potential conjugate to the angular momentum is given by
\be
	\label{eq:w:stu}
	w^{\text{STU}} = - \frac{{\cal J}}{\sqrt{\Theta^{\text{STU}} (-4\? q_0 p^1 p^2 p^3 - {\cal J}^2)}}\ .
\ee
The values of the sections $X^I$ at the south pole $\theta = 0$ can be written as
\bea
 \label{eq:X:hor:SP}
 & X^0_{\text{SP}} \equiv X^0 (\theta) \Big|_{\theta = 0} = - \frac{2 w^{\text{STU}} p^1 p^2 p^3}{\cJ} \, , \\
 & X^i_{\text{SP}} \equiv X^i (\theta) \Big|_{\theta = 0} = -w^{\text{STU}} p^i + \ii\? \frac{p^i (1+2 p^i )}{\Theta^{\text{STU}}} \, , \qquad i = 1 , 2 , 3 \, .
\eea
Their values at the north pole $\theta = \pi$ can be obtained by sending $w^{\text{STU}} \to - w^{\text{STU}}$ and $\cJ \to - \cJ$.
These become important later when we compare the entropy \eqref{STU:entropy} with the refined twisted index of $\cN=4$ SYM.

Finally, we can explicitly write down the gauge fields that complete the specification of the supersymmetric background,
\bea
A^0_{(4)} & = \frac{2\? (\Theta^{\text{STU}})^2\? p^1 p^2 p^3}{\Pi^{\text{STU}}\?  \Xi^{\text{STU}} (\theta)}\? (r\? {\rm d} t + \omega_0^{\text{STU}}) \, , \\
A^i_{(4)} & = \frac{(\Theta^{\text{STU}})^2 {\cal J}\? p^i\? \cos (\theta)}{\Pi^{\text{STU}}\?  \Xi^{\text{STU}} (\theta)}\?  (r\? {\rm d} t + \omega_0^{\text{STU}})  - p^i \cos (\theta)\? {\rm d} \phi \, ,
\eea
with the one-form
\be
	\label{eq:one-formSTU}
	\omega_0^{\text{STU}} =  \frac{{\cal J}}{\Theta^{\text{STU}}} \sin^2 (\theta)\? {\rm d} \phi \, .
\ee

\subsubsection{Dyonic case}
Adding non-vanishing electric charges $q_i$ is straightforward given our general results above. The constraint \eqref{eq:I4:id:zero} can be solved in this case by fixing one of the electric charges, \eg\;
\be
\label{eq:q3relation}
	q_3 = - \frac{q_1\? p^1\? (1+2 p^1) + q_2\? p^2\? (1+2 p^2)}{p^3\? (1+2 p^3)} \, .
\ee
The entropy is given by
\be
	\label{STU:entropydyonic}
	S^{\text{STU}}_{\text{dyonic}} = \frac{\pi}{G_{\text{N}}}\? \sqrt{\frac{-I_4^{\text{STU}}(\Gamma)- {\cal J}^2}{\Theta^{\text{STU}} } }\, ,
\ee
in terms of the quantities defined in  \eqref{I4-stu}.
The chemical potential conjugate to the angular momentum is given by
\be
	\label{eq:w:studyonic}
	w^{\text{STU}}_{\text{dyonic}} = - \frac{{\cal J}}{\sqrt{\Theta^{\text{STU}} (-I_4^{\text{STU}}(\Gamma) - {\cal J}^2)}} \, ,
\ee
while the sections, using explicitly the relation \eqref{eq:q3relation}, read
\bea
 X^0_{\text{dyonic}} (\theta) & = \frac{2\? p^1 p^2 p^3}{\sqrt{\Theta^{\text{STU}}\? \Xi^{\text{STU}}_{\text{dyonic}}(\theta)}} \, , \\
 X^i_{\text{dyonic}} (\theta) & = \frac{ p^i {\cal J} \cos (\theta)}{\sqrt{\Theta^{\text{STU}}\? \Xi^{\text{STU}}_{\text{dyonic}}(\theta)}}
 + \ii\? \frac{p^i (1+2 p^i )}{\Theta^{\text{STU}}} \\
 & + \frac{2 p^1 p^2 p^3}{(1+2 p^i) \sqrt{\Theta^{\text{STU}}\? \Xi^{\text{STU}}_{\text{dyonic}}(\theta)}} \bigg( \! q_i - \sum_{i=1}^3 q_i \! \bigg) \, , \qquad i = 1, 2 , 3 \, .
\eea

\subsection{The 5D uplift}
\label{sect:5dSTUmodel}

It is also straightforward to uplift to five dimensions the solutions described above, using the general rules \eqref{5D:4D:relation}.
In particular we can directly look at the STU model.
We  focus only on some noteworthy
features of the resulting near horizon solutions for rotating black strings, which have the standard locally AdS$_5$ asymptotics with a spatial boundary $S^1 \times S^2$, and upon a further uplift on $S^5$ become ten-dimensional solutions of type IIB supergravity.

The five-dimensional near horizon solution can be summarized as follows. The scalars $L^i$ are constant ($\theta$ independent) and coincide with the scalars in the static case as presented in \cite{Benini:2013cda}, 
		\be
			L^i = \frac{p^i\? (1+2 p^i)}{(- p^1 p^2 p^3\? \Pi^{\text{STU}})^{1/3}} \, .
		\ee
The gauge fields in the purely magnetic case are given by 
		\be
			A^i_{(5)} = - \sqrt{2} \? p^i \cos (\theta) \left( {\rm d} \phi - \frac{{\cal J}}{2\? p^1 p^2 p^3} {\rm d} y \right) ,
		\ee
where $y$ is the fifth coordinate that is topologically a non-contractible circle.
In the dyonic case the gauge fields are supplemented with additional constant Wilson lines $\alpha^i {\rm d} y$ with the constraint $\sum_i \alpha^i = 0$ as discussed in \cite{Hristov:2014hza},
where you can find the map between the charges $q_i$ and the parameters $\alpha^i$. Notice that the electric charges from a five-dimensional perspective do not alter the scalars,
meaning that the attractor mechanism is qualitatively different than its four-dimensional analogue and fixes not only the scalars, but also some components of the gauge fields.
		
The metric, as already anticipated in the introduction, is a fibration of BTZ and $S^2$ where in the limit ${\cal J} \rightarrow 0$ the fibration becomes trivial. The BTZ part is of course locally isometric to AdS$_3$, but the global difference is important in order to keep the relation to four dimensions well-defined, and, therefore, to ensure that our construction here is correct.
Note that the length scales of the BTZ and the $S^2$ do not depend on the angular momentum ${\cal J}$, which only leads to the nontrivial fibration.
Thus, the length scale associated with the two-dimensional right-moving central charge \eqref{SYM:cr} remains unchanged,
\be
	R_{\text{BTZ}}\? R^2_{S^2} = - \frac{2 p^1 p^2 p^3}{g_{(5)}^3 \Theta^{\text{STU}}} \, .
\ee

Let us finally note that the static magnetic near horizon limit of the black strings of \cite{Benini:2013cda} was already found in \cite{Kunduri:2007qy} from a classification of supersymmetric geometries.
The generalizations here with electric charges and rotation were not considered in the classification as it is crucial to allow for a globally BTZ spacetime.

\section[\texorpdfstring{$\cI$}{I}-extremization principle and microstates counting]{$\cI$-extremization principle and microstates counting}
\label{sect:holography}

We now consider the degeneracy of supersymmetric states of the $\cN=1$ gauge theores on $S^2_\omega \times T^2$ discussed in section \ref{sect:Cardy}. The chemical potentials $\Delta_a$ are conjugate to the flavor charges, $\omega$ to the angular momentum on $S^2_\omega$ and $\beta$ to the momentum along one of the cycles of the torus $T^2$. 
In a theory specified by the magnetic fluxes $\fn_a$, the number of supersymmetric  states $d_{\text{micro}}( \fn_a , e_a , J | e_0)$ with electric charges $e_a$, $a=1,\ldots, d$,  angular momentum  $J$ and momentum $e_0$   is then given by the Fourier/Laplace transform of the refined index \eqref{Cardy:formula:RTTI} with respect to the independent chemical potentials $\Delta_a$, $a=1,\ldots,d$, $\omega$, and  $\beta$,
\be
 \label{d:micro:FT}
 d_{\text{micro}} = - \frac{\ii}{(2 \pi)^{d+1}} \oint \prod_{a=1}^d \rd \Delta_a \int_{\ii \bR} \rd \beta \int  \rd \omega \? \delta \Big( 2 \pi - \sum_a \Delta_a \Big) Z ( \fn , \Delta , \omega | \beta) \? \ex^{- \ii \sum_a \Delta_a e_a - \ii \?\omega J + \beta e_0} \, .
\ee
We are considering theories of D3-branes at conical singularities and using the R-symmetry parameterization introduced in section  \ref{sect:Cardy}.
The delta function in \eqref{d:micro:FT} imposes the analogous of constraint \eqref{toric:R-charge}.  Due to this constraint, $d_{\text{micro}}$ only depends on the $d-1$ combinations $e_a-e_d$, $a=1,\ldots, d-1$,  corresponding to the flavor symmetries of the theory.  
In the Cardy limit $(\beta \to 0)$, $d_{\text{micro}}$ can be evaluated by a saddle point approximation
\be
 \log d_{\text{micro}} ( \fn_a , e_a , J | e_0) \equiv \log Z ( \fn , \Delta , \omega | \beta)- \ii \sum_a \Delta_a e_a - \ii\? \omega J + \beta e_0 \Big|_{\bar \Delta_a , \? \bar\omega , \? \bar \beta} \, ,
\ee
where $\bar \Delta_a$, $\bar\omega$, and $\bar \beta$ are the critical points of the functional
\be
 \label{I-functional}
 \cI ( \fn , \Delta , \omega | \beta) \equiv \log Z ( \fn , \Delta , \omega | \beta)- \ii \sum_a \Delta_a e_a - \ii\? \omega J + \beta e_0 \, ,
\ee
under the constraint $\sum_{a=1}^d \Delta_a = 2 \pi$, with respect to $\Delta_a$, $\omega$, and $\beta$,
\bea
 & \pd_{\Delta_a} \cI ( \fn , \Delta , \omega | \beta) \Big|_{\bar \Delta_a , \? \bar\omega , \? \bar \beta} = 0 \, , \qquad a = 1, \ldots, d-1 \, , \\
 & \pd_\omega \cI ( \fn , \Delta , \omega | \beta) \Big|_{\bar \Delta_a , \? \bar\omega , \? \bar \beta}
 = \pd_\beta \cI ( \fn , \Delta , \omega | \beta) \Big|_{\bar \Delta_a , \? \bar\omega , \? \bar \beta}
 = 0 \, .
\eea
This procedure is the so-called $\cI$\emph{-extremization} principle that has been used to give a microscopic explanation of the entropy (density) of BPS black holes (strings)
\cite{Benini:2015eyy,Benini:2016rke,Hosseini:2016cyf,Cabo-Bizet:2017jsl,Hosseini:2018qsx,Hosseini:2018uzp} in diverse dimensions.
It contains two basic pieces of information:
\begin{enumerate}
 \item
 extremizing the index unambiguously determines the exact R-symmetry of the SCFT in the IR;
 \item
 the value of the index at its critical points is the (possibly regularized) number of ground states.
\end{enumerate}
We now apply this counting to the STU black holes discussed in section \ref{sect:STU:model}. 

\subsection[The case with \texorpdfstring{$e_a=0$}{e[a]=0}]{The case with $e_a=0$}

Consider for simplicity $e_a = 0$, $a=1,\ldots, d$.
Now let us plug back the value for the refined topologically twisted index in the Cardy limit (and at finite $N$) \eqref{Cardy:formula:RTTI} into \eqref{I-functional}, obtaining
\bea
 \label{I:SYM0}
\cI ( \fn , \Delta, \omega | \beta ) = \frac{\pi^2}{6 \beta} c_l ( \fn , \Delta/\pi )
 - \frac{( 2 \omega)^2}{27 \beta} \left( 3 c(\fn) - 2 a(\fn) \right) 
   - \ii \?\omega J + \beta e_0
 \, ,
\eea
Extremizing it with respect to $\Delta_a$ sets the trial left-moving central charge $c_l (\fn , \Delta/\pi)$ to its \emph{exact} value $c^{\text{CFT}}_l(\fn)$ in the IR --- in accordance with the $c$-extremization principle \cite{Benini:2012cz,Benini:2013cda}.\footnote{The gravitational anomaly $k$ is independent of $\Delta_a$, and extremizations of $c_l$ or $c_r$ are equivalent.}  
Then, the extremization of $\cI( \fn , \Delta , \omega | \beta)$ with respect to $\beta$ and $\omega$ yields. 
\be
 \label{I-functional:critical}
 \begin{aligned}
  & \bar \beta = \pi \sqrt{\frac{c_l^{\text{CFT}}(\fn)}{6} \bigg( e_0 - \frac{27 J^2}{16 ( 3 c ( \fn ) - 2 a ( \fn ) )} \bigg)^{-1}} \, , \\
  & \bar \omega = - \frac{27 \ii}{8 (3 c ( \fn ) - 2 a ( \fn ) )} \bar \beta J \, .
 \end{aligned}
\ee
Notice that the extremal value of $\omega$ is purely imaginary thus justifying the assumptions made in the derivation of \eqref{Cardy:formula:RTTI} in section \ref{sect:Cardy}.
Substituting \eqref{I-functional:critical} back into \eqref{I-functional} we finally obtain
\be
 \label{d:micro:final}
 \log d_{\text{micro}} ( \fn , J | e_0)  = 2 \pi \sqrt{\frac{c_l^{\text{CFT}}(\fn)}{6} \bigg( e_0 - \frac{27 J^2}{16 ( 3 c(\fn)-2 a(\fn) )} \bigg)} \, .
\ee
In the case of $J=0$, this is the familiar Cardy formula for a two-dimensional CFT \cite{Cardy:1986ie}.

In order to compare \eqref{d:micro:final} with the entropy of the STU black hole \eqref{STU:entropy},
we need to go to large $N$ for which $c_l = c_r$ and $c = a$. Thus,
\be
 \label{d:micro:largeN}
 \log d_{\text{micro}} ( \fn , J | e_0)  = 2 \pi \sqrt{\frac{c_r^{\text{CFT}}(\fn)}{6} \bigg( e_0 - \frac{27 J^2}{16 a(\fn)} \bigg)} \, , \qquad N \gg 1 \, .
\ee
Plugging back the values for $c_r^{\text{CFT}}(\fn)$ and $a(\fn)$ of $\cN = 4$ SYM, see \eqref{SYM:cr} and \eqref{SYM:a(n)}, into \eqref{d:micro:largeN}, we find that
\be
 \label{d:micro:SYM}
 \log d^{\cN=4}_{\text{micro}} ( \fn , J | e_0)  = 2 \sqrt{2} \? \pi \sqrt{\frac{N^2 e_0 \fn_1 \fn_2 \fn_3 - 2 J^2}{\Theta_{\cN=4}}} \, , \qquad N \gg 1 \, ,
\ee
where $\Theta_{\cN=4}$ is given in \eqref{SYM:theta}. Moreover, the values of the critical points \eqref{I-functional:critical} read
\be
 \label{SYM:critical}
 \begin{aligned}
  & \bar \beta^{\cN=4} = \sqrt{2}\? \pi \frac{N^2 \? \fn_1 \fn_2 \fn_3}{\sqrt{\Theta_{\cN=4} ( N^2 e_0 \fn_1 \fn_2 \fn_3 - 2 J^2 )}} \, , \\
  & \bar \omega^{\cN=4} = - \frac{4 \ii}{N^2 \? \fn_1 \fn_2 \fn_3} \bar \beta^{\cN=4} J \, .
 \end{aligned}
\ee
Using the AdS$_5$/CFT$_4$ relation between gravitational and SCFT parameters at large $N$,
\be
 \label{GN:N}
 \frac{\pi}{2 g_{(5)}^3 G^{(5)}_{\text{N}}} = N^2 \, ,
\ee
where $G^{(5)}_{\text{N}} = 2 \pi G_{\text{N}}$, and upon identifying
\bea
 \label{AdS:meets:CFT}
 & \fn_a \equiv - 2 p^a \, , && a = 1, 2, 3 \, , \\
 & e_0 \equiv \frac{1}{2 \sqrt{2} G_{\text{N}}} \?q_0 \, , \quad && J \equiv - \frac{1}{2 G_{\text{N}}} \cJ \, ,
\eea
we find that the entropy of the STU black hole \eqref{STU:entropy} matches \emph{precisely} our microscopic expression for the number of ground states \eqref{d:micro:SYM}:
\be
 S^{\text{STU}} = \log d^{\cN=4}_{\text{micro}} ( \fn , J | e_0) \, .
\ee
We further observe that
\be
 \label{w:omega}
 \bar \omega^{\cN=4}  = - 2 \pi \ii \?w^{\text{STU}} \, ,
\ee
and the values of the scalars at the south pole \eqref{eq:X:hor:SP} (or north pole)  are mapped to the critical points \eqref{SYM:critical}. Explicitly, we have
\bea
 \label{sections:Delta:omega}
 & X^0_{\text{SP,} \! \text{ NP}} = - \frac{1}{\pi} \? \bar\beta^{\cN=4} \, , \\
 & X^a_{\text{SP,} \! \text{ NP}} = \frac{\ii}{2 \pi} \bigg( \bar \Delta_a^{\cN=4} \pm \frac{\bar\omega^{\cN=4}}{2} \fn_a \bigg) \, , \quad && a = 1, 2, 3 \, ,
\eea
where $\bar\Delta_a$, $a=1,2,$ are given in \eqref{SYM:bar:Delta}, $\bar \Delta_3 = 2 \pi - \bar\Delta_1 - \bar\Delta_2$ and $\pm$ refers to SP and NP, respectively.%
\footnote{Recall that $r_a = \Delta_a/\pi$ parameterize an R-symmetry of the theory.}

\subsection[Dyonic \texorpdfstring{$\cN = 4$}{N=4} super Yang-Mills]{Dyonic $\cN = 4$ super Yang-Mills}

In this section we evaluate \eqref{d:micro:FT} for $\cN = 4$ SYM keeping $e_a$, $a=1,2,3$, nonzero.
Explicitly, we write
\be
 \label{d:micro:N=4}
 d_{\text{micro}} = - \frac{\ii}{(2 \pi)^{4}} \oint \prod_{a=1}^3 \rd \Delta_a \int_{\ii \bR} \rd \beta \int \rd \omega \? \delta \Big( 2 \pi - \sum_a \Delta_a \Big)
 \ex^{\frac{\pi^2}{6 \beta} c_r ( \fn , \Delta/\pi )
 - \frac{( 2 \omega)^2}{27 \beta} a(\fn)
 - \ii \sum_{a=1}^3 \Delta_a e_a - \ii \?\omega J + \beta e_0} ,
\ee
where $c_l ( \fn , \Delta/\pi )$ and $a(\fn)$ are given in \eqref{SYM:c2d:trial} and \eqref{SYM:a(n)}, respectively.
Again, in a saddle point approximation we shall extremize
\bea
 \label{I:SYM}
 \cI ( \fn , \Delta, \omega | \beta ) = \frac{\pi^2}{6 \beta} c_r ( \fn , \Delta/\pi )
 - \frac{( 2 \omega)^2}{27 \beta} a(\fn)
 - \ii \sum_{a=1}^3 \Delta_a e_a - \ii \?\omega J + \beta e_0
 \, ,
\eea
under the constraint $\sum_{a=1}^3 \Delta_a = 2 \pi$, with respect to $\Delta_1$, $\Delta_2$, $\omega$, and $\beta$.
The critical points of the $\cI$-functional \eqref{I:SYM} read 
\bea
 \label{crit:dyonic:SYM}
 \frac{\bar \Delta_1}{2 \pi} & = \frac{\fn_1 (\fn_1 - \fn_2 - \fn_3)}{\Theta_{\cN=4}}
 - \ii \frac{\sqrt{2}\?\fn_1 \fn_2 \fn_3 \left( 2 e_1 \fn_1-e_2 (\fn_1+\fn_2-\fn_3)-e_3 (\fn_1-\fn_2+\fn_3) \right)}{\Theta_{\cN=4}^{3/2} \sqrt{\cQ \? \fn_1 \fn_2 \fn_3 - 2 J^2}} \, , \\
 \frac{\bar \Delta_2}{2 \pi} & = \frac{\fn_2 (-\fn_1+\fn_2-\fn_3)}{\Theta_{\cN=4}}
 - \ii \frac{\sqrt{2}\? \fn_1 \fn_2 \fn_3 \left( 2 e_2 \fn_2 - e_1 (\fn_1+\fn_2-\fn_3) - e_3 (-\fn_1+\fn_2+\fn_3) \right)}{\Theta_{\cN=4}^{3/2} \sqrt{\cQ\? \fn_1 \fn_2 \fn_3 - 2 J^2}} \, , \\
 \bar \beta & = \frac{\pi \sqrt{2} N^2\? \fn_1 \fn_2 \fn_3}{\sqrt{\Theta_{\cN=4} ( \cQ \? \fn_1 \fn_2 \fn_3 - 2 J^2 )}} \, , \qquad
 \bar \omega = - \frac{4 \sqrt{2} \? \ii \pi}{\sqrt{\Theta_{\cN=4} ( \cQ \? \fn_1 \fn_2 \fn_3 - 2 J^2 )}} J \, ,
\eea
where we defined
\be\label{combQ}
 \cQ \equiv e_0 N^2 - \frac{2 \left( (e_1-e_2) \left( (e_1 - e_3 ) \fn_1 - (e_2 - e_3) \fn_2 \right) + ( e_1 - e_3 ) ( e_2 - e_3 ) \fn_3 \right)}{\Theta_{\cN=4}} \, .
\ee
Notice again that the extremal value of $\omega$ is purely imaginary in agreement with the assumptions made in section \ref{sect:Cardy}.
Plugging the critical points \eqref{crit:dyonic:SYM} back into the $\cI$-functional \eqref{I:SYM} we find
\bea
 \cI \Big|_{\text{crit.}} (\fn_a , e_a, J | e_0) & = 2 \sqrt{2} \pi \sqrt{\frac{\cQ\? \fn_1 \fn_2 \fn_3 - 2 J^2}{\Theta_{\cN=4}}} \\
 & + 2 \pi \ii \frac{e_1 \fn_1 ( - \fn_1 + \fn_2 + \fn_3) + e_2 \fn_2 ( \fn_1 - \fn_2 + \fn_3 ) + e_3 \fn_3 ( \fn_1 + \fn_2 - \fn_3 )}{\Theta_{\cN=4}} \, .
\eea
Demanding the reality of the $\cI$-functional at its critical point fixes one of the charges, say $e_3$, in terms of the others. We can then write
\be
 \label{e3:constraint}
 e_3 = \frac{e_1 \fn_1 ( \fn_1 - \fn_2 - \fn_3 ) + e_2 \fn_2 ( - \fn_1 + \fn_2 - \fn_3 )}{( \fn_1 + \fn_2 - \fn_3 ) \fn_3} \, ,
\ee
and, therefore,
\be
 \label{d:micro:dyonic:SYM}
 \log d_{\text{micro}} (\fn_a , e_a, J | e_0) = 2 \sqrt{2} \pi \sqrt{\frac{\cQ\? \fn_1 \fn_2 \fn_3 - 2 J^2}{\Theta_{\cN=4}}} \, .
\ee
Quite remarkably, the constraint \eqref{e3:constraint} is \emph{precisely} the one that we obtained on the supergravity side, see \eqref{eq:q3relation}, where we identify
\be\label{eac}
 e_a = \frac{1}{2 \sqrt{2} g_{(5)} G_{\text{N}}} q_a \, , \qquad a = 1, 2, 3 \, .
\ee
It would be very interesting to give a first principle derivation of this constraint. Finally, using \eqref{GN:N}, \eqref{AdS:meets:CFT}, and \eqref{eac}, 
we find that the Bekenstein-Hawking entropy of four-dimensional rotating \emph{dyonic} black holes \eqref{STU:entropydyonic} matches \eqref{d:micro:dyonic:SYM}:
\be
 S^{\text{STU}}_{\text{dyonic}} = \log d_{\text{micro}} ( \fn_a , e_a , J | e_0) \, .
\ee
Notice that the expression for the entropy matches only  after imposing the constraint \eqref{e3:constraint}.%
\footnote{In particular, $\cQ\? \fn_1 \fn_2 \fn_3 = - 2 N^4 I_4^{\text{STU}}(\Gamma)/\pi^2$ holds only if the charges satisfy \eqref{e3:constraint}.}
We observe that the relations \eqref{w:omega} and \eqref{sections:Delta:omega} hold also in this case.

It is interesting to rewrite the combination \eqref{combQ} in terms of purely field theoretic data, in particular 't Hooft flavor anomalies.
The chemical potentials $\Delta_a$  are conjugated to a basis $R_a$, $a=1,2,3$, of R-symmetries that assign charge $2$ to the $a$-th chiral multiplet of $\cN=4$ SYM and zero to the others.
We can choose two independent flavor symmetries $Q_1=(R_1-R_3)/2$ and $Q_2 =(R_2-R_3)/2$. The 't Hooft anomaly matrix   is easily computed from the multiplicity of fermionic zero-modes\footnote{Recall that  the difference between the number of fermionic zero-modes  of opposite chiralities  is equal to  $ - \text{ dim}\,\fR_I ( \fn_I - 1 )$ for four-dimensional matter fields in a representation $\fR_I$.} and  reads
\be
 A_{AB} = \Tr \gamma_3 Q_A Q_B
 = N^2 \begin{pmatrix}
 2 - \fn_1 - \fn_3 & 1 - \fn_3 \\  1 - \fn_3 & 2 - \fn_2 - \fn_3
 \end{pmatrix} \, .
\ee   
We can then write \eqref{combQ}  as
\be
 \label{modular}
 \frac{\cQ}{N^2} = e_0 + \frac{1}{2} \sum_{A,B=1}^2 \tilde e_A \? A^{-1}_{AB} \? \tilde e_B \, ,
\ee
where $\tilde e_A= e_A - e_3$ are the flavor charges. We see that the entropy depends only on a  particular combination of $e_0$ and electric flavor charges. As we will see, this result extends to more general quivers.
The combination \eqref{modular} is familiar from the Rademacher expansion for asymptotically flat black holes \cite{Dijkgraaf:2000fq,Manschot:2007ha},
where it is a consequence of the transformation properties of the elliptic genus of the two-dimensional CFT.  
Our topologically twisted index is expected to compute the elliptic genus of the IR CFT and relate to the quantum entropy of black holes not just in the Cardy limit. The result, at finite $N$, is actually a  meromorphic function of $\Delta_a$,
and has a complicated structure in terms of modular forms. We expect that, for large $N$, this structure simplifies and gives results similar to those
for asymptotically flat black holes. In the latter case the meromorphicity of the elliptic genus leads to interesting behavior, related to wall-crossing phenomena (see for example \cite{Dabholkar:2012nd}). It would be interesting to see if the same happens for the black holes discussed here once quantum corrections are taken into account.

\subsection{Microstate counting for a generic quiver}\label{sec:genquiver}

Although there is no gravitational dual at the moment, for completeness we count the degeneracy of states for a generic theory of D3-branes at toric conical singularities.

We must extremize the analog of \eqref{I:SYM},
\bea
 \label{I:SYM2}
 \cI ( \fn , \Delta, \omega | \beta ) = \frac{\pi^2}{6 \beta} c_r ( \fn , \Delta/\pi )
 - \frac{( 2 \omega)^2}{27 \beta} a(\fn)
 - \ii \sum_{a=1}^d \Delta_a e_a - \ii \?\omega J + \beta e_0
 \, ,
\eea
with respect to $\omega$, $\beta$ and $d$ chemical potential $\Delta_a$ subject to the constraint $\sum_{a=1}^d \Delta_a = 2 \pi$.
It is convenient to introduce a basis of flavor symmetries $Q_A=(R_A-R_d)/2$, $A=1,\ldots, d-1,$
where $R_a$ is the basis of R-symmetries conjugated to the chemical potentials $\Delta_a$.
Choosing a reference R-symmetry $r_a^0$, we can parameterize
\bea
 & \frac{\Delta_a}{\pi} = r_a^0 +\delta_a \, , \qquad a = 1, \ldots, d \, , \\
 & \sum_{a=1}^d r_a^0=2\, ,\qquad \sum_{a=1}^d \delta_a = 0 \, ,
\eea
so that the two-dimensional R-symmetry matrix reads
\be
 R(\Delta_a/\pi) = R_0 + \sum_{A=1}^{d-1} \delta_A Q_A \, , \qquad R_0\equiv R(r^0_a) \, ,
\ee 
and the two-dimensional trial central charge $c_r$ becomes
\be
 3 \Tr \gamma_3 R(\Delta_a/\pi)^2 = 3 \Tr \gamma_3 R_0^2 + 3 \sum_{A,B=1}^{d-1} \delta_A A_{AB} \delta_B + 6 \sum_{A=1}^{d-1} k_A \delta_A \, ,
\ee
where
\be
 A_{AB}= \Tr \gamma_3 Q_A Q_B\, ,\qquad k_A = \Tr  \gamma_3 R_0 Q_A \, .
\ee
Notice that $A_{AB}$ is the 't Hooft anomaly matrix and is a function of the fluxes $\fn_a$. Plugging it back into \eqref{I:SYM2}, and extremizing with respect to $\beta$, $\omega$ and the $d-1$ independent parameters $\delta_A$
we find
\bea
 \bar \beta &= \pi \sqrt{\frac{\Tr \gamma_3 R_0^2 -  k_A A_{AB}^{-1} k_B }{ 2 ( e_0 +\frac12  \tilde e_A A_{AB}^{-1} \tilde e_B - \frac{27}{16 a(\fn)} J^2) }}
 = \pi \sqrt{\frac{c_r^{\text{CFT}}(\fn)}{ 6 ( e_0 +\frac12  \tilde e_A A_{AB}^{-1} \tilde e_B - \frac{27}{16 a(\fn)} J^2) }} \, ,\\
 \bar \omega & = - \frac{27 \ii \bar \beta}{8 a(\fn)} J \, , \qquad\qquad  \bar \delta_B  = \ii A^{-1}_{AB} \left (  \frac{ \bar \beta}{\pi}  \tilde e_A + \ii k_A\right ) \, , \qquad   A, B =1,\ldots , d-1 \, ,
\eea 
where $\tilde e_A = e_A - e_d$ and, here and below, summation over repeated indices is understood.  In the first line we reconstructed the exact central charge $c_r^{\text{CFT}}(\fn)$
by observing that the result must reduce to \eqref{I-functional:critical} for $e_a=0$. We also find the critical value of the functional \eqref{I:SYM2}
\bea\label{extI}
 \cI \Big|_{\text{crit.}} (\fn_a , e_a, J | e_0) = 2  \pi \sqrt{ \frac{c_r^{\text{CFT}}(\fn)}{6} \left( e_0 +\frac12  \tilde e_A A_{AB}^{-1} \tilde e_B - \frac{27}{16 a(\fn)} J^2\right ) } + \pi \ii \left ( \tilde e_A A_{AB}^{-1} k_B - r_a^0 e_a \right ) .
\eea
By identifying the degeneracy of states with the real part of the $\cI$-functional, we finally find
\bea
 \re \log d_{\text{micro}} (\fn_a , e_a, J | e_0)  = 2  \pi \sqrt{ \frac{c_r^{\text{CFT}}(\fn)}{6} \left( e_0 +\frac12  \tilde e_A A_{AB}^{-1}(\fn)  \tilde e_B - \frac{27}{16 a(\fn)} J^2\right ) }  \, .
\eea
We see that the field theory entropy can be completely written in terms of anomalies and central charges of the four- and two-dimensional theories.
It also depends on a very particular combinations of electric charges and $J$. It would be interesting to see if, in the dual black hole, the imaginary part of \eqref{extI} vanishes 
as a consequence of the gravitational BPS constraint on electric charges \eqref{eq:I4:id:zero}, as it happens for $\cN=4$ SYM. 

\section{Discussion and outlook}
\label{sect:discussion}

In this paper we have studied a class of rotating black strings that can be embedded in AdS$_5\times S^5$. In five-dimensional language, these are domain walls  that interpolate between AdS$_5$ and a near horizon region consisting of a warped  fibration of BTZ over a sphere, thus generalizing the solutions  of \cite{Benini:2013cda} by adding rotation. Upon compactification along the BTZ circle, we find a rotating black hole with hvLif asymptotics. We have successfully matched the entropy of such black holes with a field theory computation based on the refined topologically twisted index of $\cN=4$ SYM on $T^2\times S^2_\omega$ in the Cardy limit.  We have also computed the index in the Cardy limit at finite $N$ for a generic $\cN=1$ SCFT living on the world-volume of D3-branes at toric (and not only) conical singularities.

Many interesting questions remain open. It would be very interesting to generalize the gravity solutions that we have found in this paper to a generic compactification based on Sasaki-Einstein manifolds and compare the entropy of the resulting four-dimensional black holes with the field theory prediction of section  \ref{sec:genquiver}. There are various obstacles to do that. First of all, consistent truncations to a five-dimensional gauged supergravity contaning only vector multiplets exist just for AdS$_5\times S^5$. The available truncations on AdS$_5\times {\rm SE}_5$ \cite{Gauntlett:2007ma,Cassani:2010na,Bena:2010pr}, where ${\rm SE}_5$ is a five-dimensional Sasaki-Einstein manifold, involve hypermultiplets as well and are more difficult to treat. Such truncations typically contain massive vector multiplets and the hypermultiplets themselves have a vacuum expectation value that Higgses other vectors. Therefore, the number of vector multiplets in the truncation needs not to be equal  to  the number of global symmetries of the dual SCFT, since the latter correspond to massless vectors in the bulk. It would be tempting to speculate that, as far as the near horizon region is concerned, we should be able to use an effective supergravity containing only massless vector multiplets associated with the global symmetries. This happens, for example, in the matching of the entropy of some magnetically charged AdS$_4$ black holes \cite{Hosseini:2017fjo,Benini:2017oxt,Bobev:2018uxk}. Since a five-dimensional gauged supergravity with only vector  multiplets  is completely specified by the tensor $c_{ijk}$, associated with the 't Hooft anomalies of the dual CFT, the entropy will be also a function of $c_{ijk}$, $J$ and the magnetic and electric charges. Indeed we found that the field theory prediction for the entropy given  in section  \ref{sec:genquiver} can be written just in terms of anomalies. However, we cannot immediately compare with our supergravity results. 
The solutions found in this paper assume the validity of \eqref{eq:symm-cub}, which, unfortunately, is not satisfied by the anomaly coefficients of a generic $\cN=1$ SCFT of D3-branes at singularities.
From this perspective, it would be very interesting to remove the technical assumption \eqref{eq:symm-cub}.

It would be also very interesting to see what part of our analysis can be extended to the refined topologically twisted index on $S^1\times S^2_\omega$, which is much more difficult to analyse from the field theory point of view.  The refined topologically twisted index in three dimensions is supposed to count the microstates of the magnetically charged and rotating asymptotically AdS$_4$ black holes found in \cite{Hristov:2018spe}.  We hope to report soon on the subject.
  
\section*{Acknowledgements}

KH would like to thank Stefanos Katmadas for countless explanations about black holes and the quartic invariant over the years.
The work of SMH was supported by World Premier International Research Center Initiative (WPI Initiative), MEXT, Japan.
KH is supported in part by the Bulgarian NSF grants DN08/3 and N28/5.
AZ is partially supported by the INFN, the ERC-STG grant 637844-HBQFTNCER and the  MIUR-PRIN contract 2017CC72MK003.
SMH would like to thank the Institute for Theoretical Physics at KU Leuven for their kind hospitality during his visit, where part of this work was done.

\begin{appendix}

\section[Asymptotic behavior of elliptic functions near \texorpdfstring{$q = 1$}{q=1}]{Asymptotic behavior of elliptic functions near $q = 1$}
\label{app:special:functions}

The Dedekind eta function is defined by
\be
 \label{dedekind:eta}
 \eta(q) = \eta(\tau) = q^{\frac{1}{24}} \prod_{n=1}^{\infty} \left(1 - q^{n} \right) \, , \qquad \qquad \im \tau > 0 \, ,
\ee
and it has the following modular properties
\be
 \label{eta:modular}
\eta(\tau + 1) = \ex^{\frac{\ii \pi }{12}} \, \eta (\tau)\, , \qquad \eta \Big( \! \! - \frac{1}{\tau} \Big) = \sqrt{- \ii \tau} \, \eta (\tau) \, .
\ee
Here, $q = \ex^{2 \pi \ii \tau}$.
The Jacobi theta function is also defined by
\bea
 \label{theta:function}
 \theta_1 (x;q) = \theta_1 (u;\tau) & = - \ii q^{\frac18} x^{\frac12} \prod_{k=1}^{\infty} \left( 1- q^k \right) \left( 1-x q^k \right) \left( 1- x^{-1} q^{k-1} \right) \\
 & = - \ii \sum_{n \in \mathbb{Z}} (-1)^n \ex^{\ii u \left( n+ \frac12 \right)} \ex^{\pi \ii \tau \left( n+ \frac12 \right)^2}\, ,
\eea
where $x = \ex^{\ii u}$ and $q$ is as before.
\eqref{theta:function} has simple zeros in $u$ at $u = 2 \pi \mathbb{Z} + 2 \pi \tau \mathbb{Z}$ and no poles.
Its modular properties read
\be
 \label{theta:modular}
 \theta_1 (u;\tau+1 ) = \ex^{\frac{\ii \pi}{4}} \, \theta_1 (u;\tau) \, , \qquad
 \theta_1 \Big( \frac{u}{\tau};-\frac{1}{\tau} \Big) = - \ii \sqrt{- \ii \tau} \, \ex^{\frac{\ii u^2}{4 \pi \tau}} \, \theta_1 ( u; \tau ) \, .
\ee
Furthermore, \eqref{theta:function} satisfies the following useful formula
\be
 \label{theta:shift}
 \theta_1\left( q^m x;q\right) = (-1)^{-m}\,x^{-m} q^{-\frac{m^2}{2}}\theta_1(x;q) \, , \qquad m \in \bZ \, .
\ee

The asymptotic behavior of \eqref{dedekind:eta} and \eqref{theta:function} near $q = 1$ $(\tau = \ii 0)$ can be easily derived by using their modular properties \eqref{eta:modular} and \eqref{theta:modular}, respectively.
This was already done in \cite[app.\,A]{Hosseini:2016cyf} and here we only quote the final result, \ie\;
\bea
 \label{eta:theta:near:q=1}
 & \log \eta( \beta) = - \frac{1}{2}\log \Big( \frac{\beta}{2\pi} \Big)-\frac{\pi^2}{6\beta} + \cO\left( \ex^{- 1 / \beta} \right) \, , \\
 & \log \theta_1(u;\beta)  = -\frac{\pi^2}{2\beta} - \frac{u^2}{2\beta}-\frac{1}{2} \log \Big( \frac{\beta}{2\pi} \Big)
+ \frac{\pi}{\beta} u \sign\left[ \re(u) \right] + \log \left( \sign \left[ \re(u) \right] \right)
+ \cO \left( \ex^{- 1 / \beta} \right) \, .
\eea
Here we identified the complex structure of the torus $\tau$ with the \emph{fictitious} inverse temperature $\beta$ as $\tau \equiv \ii \beta / ( 2 \pi )$.
Therefore, for $\re( u )> 0$,
\be
 \label{1-loop:near:q=1}
 \log \left( \frac{\ii \eta (q)}{\theta_1 ( x ; q)} \right) \sim \frac{1}{\beta} g_2 ( u ) + \frac{\ii \pi }{2} \, , \quad \text{ as } \beta \to 0 \, ,
\ee
where we defined the polynomial functions
\be
 \label{g:functions}
 \begin{aligned}
   g_2 ( u ) = \frac{u^2}{2} - \pi u + \frac{\pi^2}{3} \, , \qquad
   g_3 ( u ) = \frac{u^3}{6} - \frac{\pi}{2} u^2 + \frac{\pi^2}{3} u \, .
 \end{aligned}
\ee

\end{appendix}

\bibliographystyle{ytphys}

\bibliography{4DRTTI-HHZ}

\end{document}